\documentclass[12pt]{article}
\usepackage{amssymb}
\usepackage{graphicx}
\usepackage{amsmath,amscd,enumerate}
\def\be{\begin{eqnarray}}
\def\ben{\begin{eqnarray*}}
\def\ee{\end{eqnarray}}
\def\een{\end{eqnarray*}}
\def\Tr{{\rm Tr}}

\def\D{\mathcal{D}}

\def\=:{=\hspace{-.7em}\raisebox{1.1ex}{.}\hspace{.1em}\raisebox{-0.2ex}{.} }
\newcommand{\NF}{N_{\rm F}}
\newcommand{\NC}{N_{\rm C}}

\def\N{{\cal N}}

\newcommand {\beq}{\begin{eqnarray}}
\newcommand {\eeq}{\end{eqnarray}}

\makeatletter
\newcommand{\thetablename}{Table}
\def\fnum@table{\thetablename\ \thetable}
\makeatother
\begin{document}
\thispagestyle{empty}
\begin{flushright}
TIT/HEP--544 \\
{\tt hep-th/0508241} \\
September, 2005 \\
\end{flushright}
\vspace{3mm}

\begin{center}
{\Large \bf 
Non-Abelian Webs of Walls
}
\\[12mm]
\vspace{5mm}

\normalsize
{\large %\bf 
Minoru~Eto}\!\!
\footnote{\it  e-mail address: 
meto@th.phys.titech.ac.jp
}, 
  {\large %\bf 
Youichi~Isozumi}\!\!
\footnote{\it  e-mail address: 
isozumi@th.phys.titech.ac.jp
}, 
  {\large %\bf 
Muneto~Nitta}\!\!
\footnote{\it  e-mail address: 
nitta@th.phys.titech.ac.jp
}, \\
  {\large %\bf 
 Keisuke~Ohashi}\!\!
\footnote{\it  e-mail address: 
keisuke@th.phys.titech.ac.jp
} 
~and~~  {\large %\bf 
Norisuke~Sakai}\!\!
\footnote{\it  e-mail address: 
nsakai@th.phys.titech.ac.jp
} 

\vskip 1.5em

{ \it Department of Physics, Tokyo Institute of 
Technology \\
Tokyo 152-8551, JAPAN  
 }
\vspace{12mm}

%{\bf Abstract}\\[5mm]
%{\parbox{13cm}{\hspace{5mm}
%%%%%%%%%%%%%%%%%%%%%%%%%%%%%%%%%%%%%%%%%%%%%%%%%%
\abstract{
Domain wall junctions are studied in ${\cal N}=2$ 
supersymmetric $U(N_{\rm C})$
gauge theory with $N_{\rm F}(>N_{\rm C})$ flavors. 
We find that all three possibilities are 
realized for positive, negative and zero junction charges.
The positive junction charge is found 
to be carried by a topological charge in the Hitchin system 
of an $SU(2)$ gauge subgroup.
We establish rules of the construction of the webs of walls.
Webs can be understood qualitatively by grid diagram 
and quantitatively by associating moduli 
parameters to web configurations. 
}
%%%%%%%%%%%%%%%%%%%%%%%%%%%%%%%%%%%%%%%%%%%%%%%%%%
%}}
\end{center}
\vfill
\newpage
\setcounter{page}{1}
\setcounter{footnote}{0}
\renewcommand{\thefootnote}{\arabic{footnote}}

%%%%%%%%%%%%%%%%%%%%%%%%%%%%%%%%%%%%%%%%%%%%%%%%%%%%%%%%%%%%%%%%%%%%%%%%%%%%

%%%%%%%%%%%% text start %%%%%%%%%%%%%%%%%%%

\subsubsection*{Introduction and Summary}
The domain wall is one of the important solitons in many 
areas of physics, such as particle physics, cosmology and 
condensed matter physics. 
Recently, the 1/2 BPS domain walls in the ${\cal N}=2$ 
supersymmetric (SUSY) non-Abelian $U(\NC)$ 
gauge theories were intensively studied 
and their moduli space was found to be the
complex Grassmann manifold \cite{Isozumi:2004jc,Eto:2004vy}. 
When several non-parallel 1/2 BPS domain walls coexist,
a 1/4 of SUSY is preserved by the configuration, which is 
called a 1/4 BPS state \cite{Gibbons:1999np}.
In the previous paper \cite{Eto:2005cp} 
we have found that multiple walls in the Abelian-Higgs 
system develop a web as a 1/4 BPS state 
similarly to
$(p,q)$-string/5-brane webs in superstring theory. 
The junction charge, called the $Y$-charge, 
has been found to be always negative in the Abelian 
gauge theory \cite{Eto:2005cp, Kakimoto:2003zu} as well as 
in generalized Wess-Zumino models \cite{Oda:1999az}, 
which can be understood as 
binding energy of the constituent domain walls.
We also have found that the total moduli space of the webs, 
defined by all topological sectors patched together,  
to be the complex Grassmann manifold for the $U(N_{\rm C})$ 
gauge theory.

The purpose of this paper is to study 
1/4 BPS webs of walls in the non-Abelian gauge theories 
by extending the previous analysis 
\cite{Eto:2005cp} 
to the ${\cal N}=2$ SUSY $U(\NC)$ gauge theories 
coupled with $\NF(>N_{\rm C})$ Higgs fields (hypermultiplets) 
in the fundamental representation. 
There exist two kinds of  {nontrivial} domain wall junctions. 
One is the {\it Abelian junction} 
characterized by a negative $Y$-charge. 
The other is the {\it non-Abelian junction} 
with a positive $Y$-charge. 
We find that opposite sign of the $Y$-charge can be 
attributed to quite a different 
internal structures of the Abelian and non-Abelian 
junctions. 
The non-Abelian $Y$-charge can be identified 
with the charge of the Hitchin system
 \cite{Eto:2005sw} 
which is positive. 
Besides these two kinds of junctions, 
there exist trivial intersections of 
penetrable walls \cite{Isozumi:2004jc} with vanishing 
$Y$-charge. 
Generally, walls in the non-Abelian gauge theories 
constitute very rich variety of 
webs of the Abelian and non-Abelian junctions and the 
intersections of penetrable walls. 
We find rules of the construction of the webs.
Qualitative properties of the webs can be 
easily understood in terms of the 
grid diagram capturing the most relevant informations 
of the complex Grassmann manifold. 
On the other hand, their quantitative properties are 
clarified by moduli parameters corresponding
to web configurations. 
Especially, difference between the Abelian junctions and 
the non-Abelian junctions can be understood by an 
embedding relation of the complex Grassmann manifold 
into a complex projective space which is called 
Pl\"ucker embedding.  
We explicitly identify 
normalizable modes of the webs 
with loops, which gives 
a $1+1$ dimensional 
${\cal N}=(2,0)$ SUSY sigma model as the effective 
theory on the web.

This paper is organized by two parts. 
We devote the first part to clarify 
the qualitative properties of the webs. 
In the second part we concentrate on their 
quantitative properties. 
\vspace*{-.3cm}

\subsubsection*{Webs of walls in the SUSY 
Yang-Mills theory}
\vspace*{-.3cm}

Let us start with the 1+3 dimensional $\N = 2$ 
SUSY $U(\NC)$ gauge theory coupled 
to the $\NF(>\NC)$ hypermultiplets 
in the fundamental representation. 
The physical bosonic fields are a gauge field 
$W_\mu$ and a complex scalar $\Sigma = \Sigma_1+i\Sigma_2$, 
in the vector multiplet and complex scalars $H^{irA}$ 
$(r=1,2,\cdots,\NC,\ A=1,2,\cdots,\NF,\ i=1,2)$
in the hypermultiplets. 
We turn on completely non-degenerate complex masses 
for the hypermultiplets, which can be expressed as a 
diagonal mass matrix $M={\rm diag}(\mu_1,\cdots,\mu_{\NF})$ 
with $\mu_A \in {\bf C}$. 
In the following we will use both
the complex notation $\mu_A = m_A + in_A\ (M=M_1+iM_2)$ and 
the two-vector notation $\vec\mu_A = (m_A,n_A)$ for masses.
We also turn on the Fayet-Iliopoulos (FI) parameter 
$c>0$ in the third direction
of $SU(2)_R$ triplet.
We consider minimal kinetic terms for all the fields. 
The scalar potential is 
\be
V = \Tr\left[\frac{1}{g^2}\sum_{a=1}^{3}\left(Y^a\right)^2
+ \sum_{\alpha=1}^2\left(H^i M_\alpha - \Sigma_\alpha H^i\right)
\left(H^i M_\alpha - \Sigma_\alpha H^i\right)^\dagger
- \frac{1}{g^2}\left[\Sigma_1,\Sigma_2\right]^2
\right],
\ee
with $
Y^a \equiv \frac{g^2}{2}
\left(c^a{\bf 1}_{\NC} - (\sigma^a)^j{_i}H^i(H^j)^\dagger\right)
$. 
There exist $_{\NF}C_{\NC} = \NF!/\NC! (\NF-\NC)!$ 
discrete vacua where gauge symmetry is fully broken 
\cite{Arai:2003tc}.
Each vacuum is characterized 
by a set of $\NC$ different flavor indices 
$\langle A_1A_2\cdots A_{\NC}\rangle $
out of $N_{\rm F}$ flavors. 
In the rest of this article we may use a notation 
$\langle A_r\rangle$
short for $\langle A_1A_2\cdots A_{\NC}\rangle $.
In the vacuum $\langle A_r\rangle$,  
the vacuum expectation values (VEVs) of the hypermultiplets 
are $H^{1rA} = \sqrt c \delta^{A_r}{_A},\ H^{2rA} = 0$
and the VEV of the adjoint scalar is
$
\Sigma  
= {\rm diag}(\mu_{A_1},\cdots,\mu_{A_{\NC}}).
$
We express $\NC\times\NF$ matrix of the hypermultiplets 
by $H^i$ and set $H^2=0$ with $H\equiv H^1$ 
when we consider domain wall solutions
in the following.

Single 1/2 BPS walls 
interpolate between two vacua 
$\left<\ \underline{\cdots}\ A\right>$
and $\left<\ \underline{\cdots}\ B\right>$ 
with only one different label.
Here the underlines denote  
the same set of integers (ordering of integers does 
not affect anything).
These walls are characterized 
by a complex central charge 
in the SUSY algebra
\be
Z = c(\mu_A - \mu_B) = c\left[(m_A-m_B)+i(n_A-n_B)\right].
\label{central_charge}
\ee
By reexpressing this charge as $Z = T e^{i\theta}$, 
we can read the tension $T$ 
and the angle $\theta$ of normal vector 
$\vec n_\theta=(\cos\theta,\sin\theta)$ 
to the wall in the spatial $x^1$-$x^2$ plane.

\begin{figure}[ht]
\begin{center}
\begin{tabular}{ccc}
\includegraphics[height=4cm]{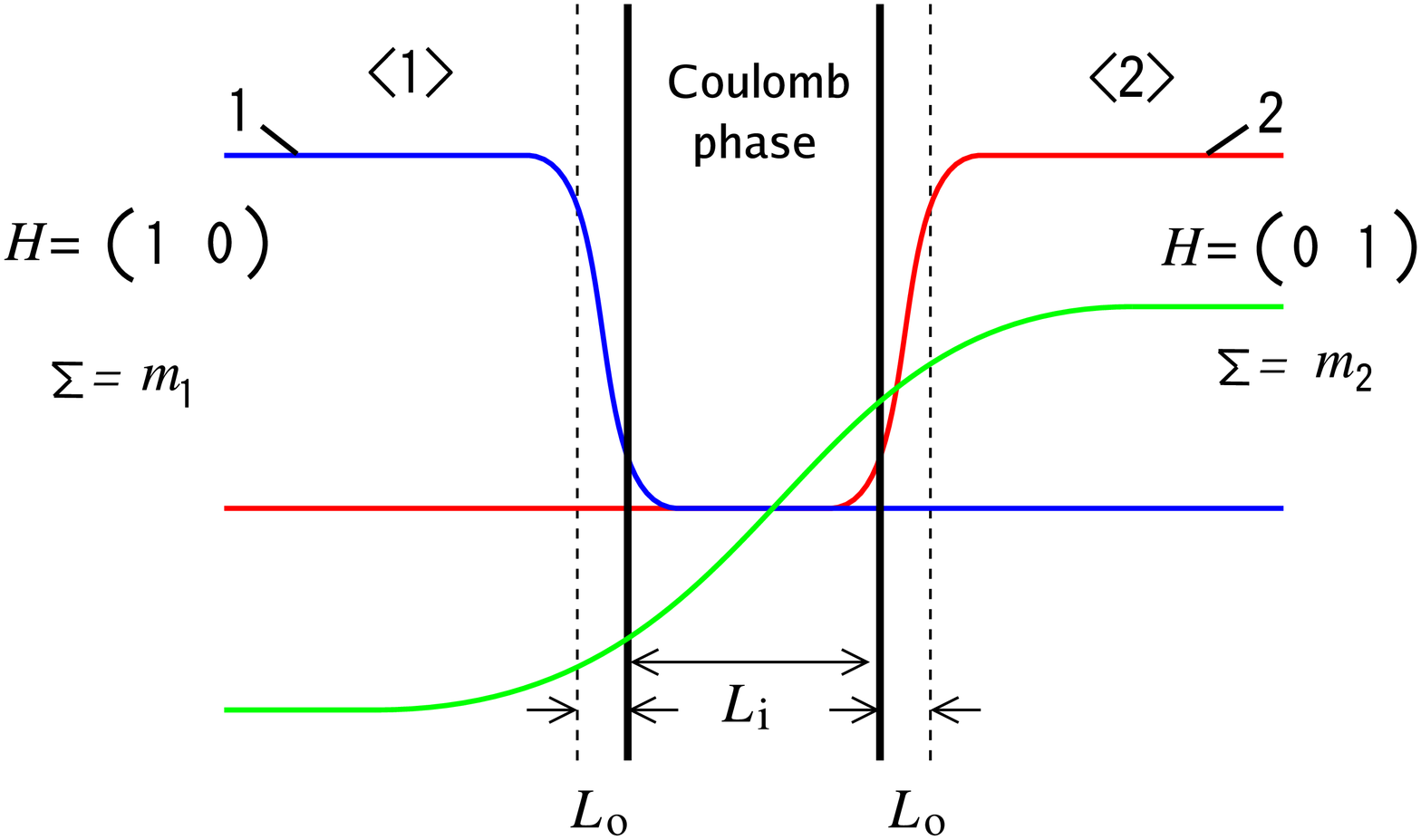}
&\quad&
\includegraphics[height=4cm]{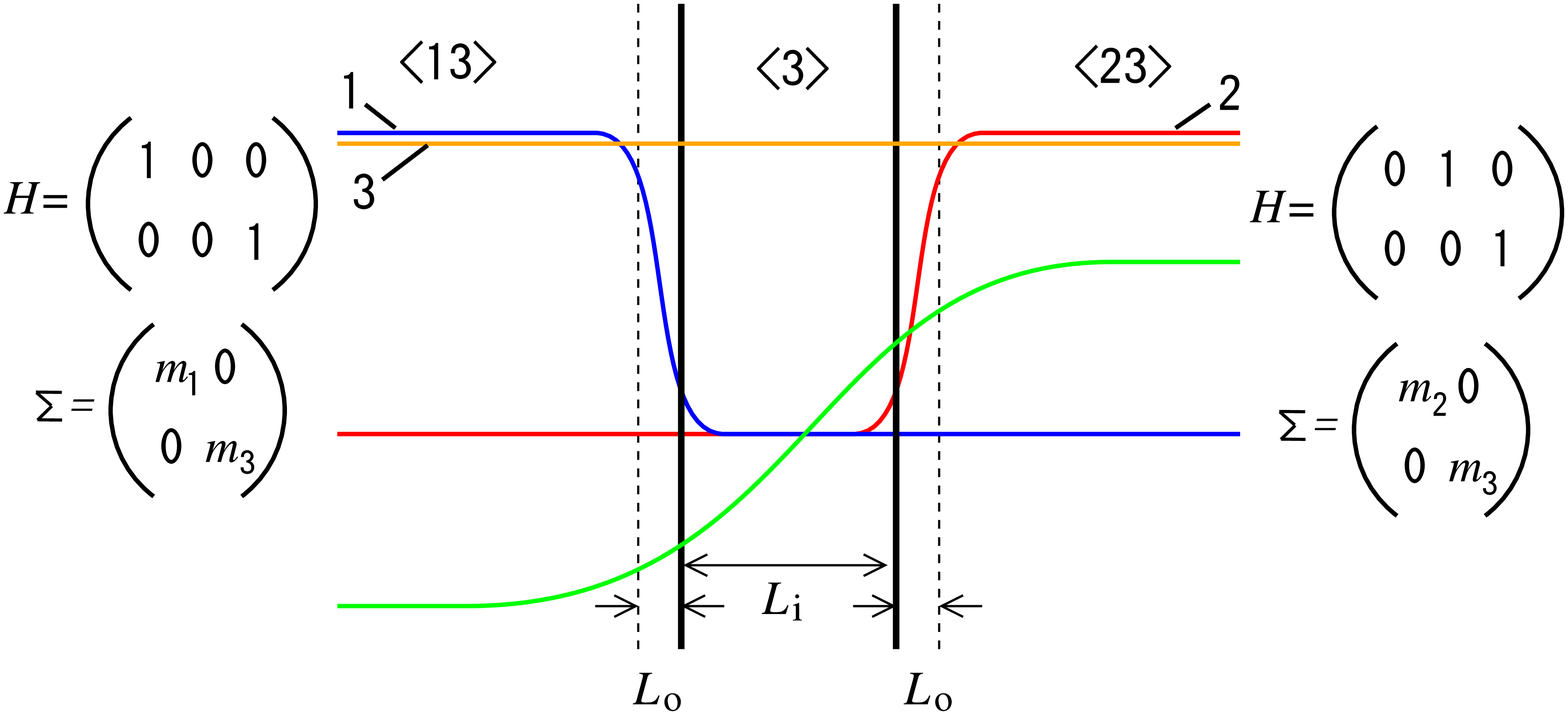}\\
\small{\sf (a) Abelian gauge theory} & & 
\small{\sf (b) non-Abelian gauge theory}
\vspace*{-.3cm}
\end{tabular}
\caption{\small{\sf Internal structures of the domain walls 
with $g\sqrt c \ll |\Delta m+i\Delta n|$.}}
\label{wall}
\vspace*{-.3cm}
\end{center}
\end{figure}

Domain walls in the Abelian gauge theory were known to 
have the following internal structures \cite{Shifman:2002jm}.
There are two cases according to the value of the 
dimensionless parameter $g\sqrt c/|\Delta m+i\Delta n|$.
Walls have a three-layer structure shown in 
Fig.~\ref{wall}(a), 
in the case of $g \sqrt c \ll |\Delta m+i\Delta n|$ 
(at weak gauge coupling).
The outer two thin layers have the same width of 
order $L_{\rm o}=1/g\sqrt c$ 
and the internal fat layer has width of order 
$L_{\rm i}=|\Delta m+i\Delta n|/g^2c$.
In the Fig.~\ref{wall}(a) the wall interpolates between 
the vacuum 
$\langle 1\rangle\ \left(H=\sqrt{c}(1,0),\ \Sigma = m_1\right)$ 
at $x^1\to -\infty$ and the vacuum 
$\langle 2\rangle\ \left(H=\sqrt{c}(0,1),\ \Sigma = m_2\right)$ 
at $x^1\to +\infty$.
The first (second) flavor component of the Higgs field 
exponentially decreases 
in the left (right) outer layer 
so that the entire $U(1)$ gauge symmetry is restored in 
the inner core. 
On the other hand, in the case of the walls in the 
non-Abelian gauge theories, 
the internal structure is a little bit different 
from that in the Abelian gauge theory. 
For simplicity, let us explain it in the $U(2)$ gauge theory
with three flavors. 
A wall interpolating between the vacuum 
$
\langle 13\rangle\ 
\big( H = \sqrt c
\left(
\begin{smallmatrix}
1 & 0 & 0\\
0 & 0 & 1
\end{smallmatrix}
\right),\ 
\Sigma =
\left(
\begin{smallmatrix}
m_1 & 0\\
0 & m_3
\end{smallmatrix}
\right)\big)
$
at $x^1\to-\infty$ and the vacuum 
$\langle 23\rangle\ 
\big(
H= \sqrt c \left(
\begin{smallmatrix}
0 & 1 & 0\\
0 & 0 & 1
\end{smallmatrix}
\right),\ 
\Sigma =
\left(
\begin{smallmatrix}
m_2 & 0\\
0 & m_3
\end{smallmatrix}
\right)\big)
$
at $x^1\to +\infty$ has the 
three-layer structure in the limit 
$g\sqrt c \ll |\Delta m+i\Delta n|$ 
(at weak gauge coupling). 
The first and the second components of the Higgs fields 
exponentially decrease in the two outer layers 
and almost vanish in the middle layer, whereas 
the third flavor component gets non-vanishing 
values over the whole region. 
Hence only the Abelian subgroup acting on 
the first color component is recovered in the middle layer 
of the wall, while 
the overall $U(1)$ is broken. 
We denote this phase in the middle layer by 
the flavor with a constant VEV like $\langle 3\rangle$ as 
shown in Fig.~\ref{wall}(b).
In the opposite limit $g\sqrt c \gg |\Delta m+i\Delta n|$ 
(at strong gauge coupling),
the internal structure becomes simpler
for both Abelian and non-Abelian cases.\footnote{
In the strong gauge coupling limit the model
becomes a nonlinear sigma model whose target space is $T^\star
G_{\NF,\NC}$~\cite{Arai:2003tc}.
In this limit the BPS equation for 
1/4 BPS wall junctions can be exactly solved \cite{Eto:2005cp}.
}
There the middle layer 
disappears 
while two outer layers of the Higgs phase grow.
The width of the wall is of order $1/|\Delta m+i\Delta n|$.

A web of walls contains several constituent walls with 
different slopes in the $x^1$-$x^2$ space.
Each of them preserves different 1/2 SUSY 
and their webs preserve the 1/4 SUSY.
The corresponding 1/4 BPS equations for the webs of walls 
are obtained \cite{Eto:2005cp} as 
\be
\left[\D_1+\Sigma_1,\D_2+\Sigma_2\right]=0,\quad
\D_\alpha H = HM_\alpha - \Sigma_\alpha H,\quad
\sum_\alpha\D_\alpha\Sigma_\alpha =
\frac{g^2}{2}\left(c{\bf 1}_{\NC} - HH^\dagger\right),
\label{bps_eq}
\ee
with $\alpha=1,2$. 
The 1/4 BPS condition assures the 
equilibrium for tensions of walls, and 
the conservation of central charges ($\sum_l Z_l = 0$) 
at every junction point. 
Here index $l$ labels the walls extending from that 
junction point.
Configurations of wall junctions are characterized 
by another central charge\footnote{
The $Y$-charge for junctions was previously 
considered in the Wess-Zumino model \cite{Gibbons:1999np} 
and in the ${\cal N}=1$ gauge theories \cite{Gorsky:1999hk}, 
although it was considered for another situation with axial 
symmetry earlier \cite{Chibisov:1997rc}.} ($Y$-charge) in the SUSY algebra
\be
Y = - \frac{2}{g^2} \int dx^1dx^2\ 
\partial_\alpha\Tr\left(\epsilon^{\alpha\beta}
\Sigma_1\D_\beta\Sigma_2\right).
\label{Y-charge}
\ee

In our theory elementary wall junctions are 3-pronged 
junctions where three constituent walls meet.
In the non-Abelian gauge theory
there exist two kinds of 
nontrivial junctions of domain walls
according to sets of three vacua 
divided by three walls in the junctions:
\begin{itemize}
\item {\it Abelian junction}:
Abelian junctions divide a set of three vacua 
$\left<\ \underline{\cdots}\ A\right>$,
$\left<\ \underline{\cdots}\ B\right>$ and
$\left<\ \underline{\cdots}\ C\right>$ 
with different labels in only one
color component.
This junction exists both in Abelian 
and non-Abelian gauge theories.
\item {\it Non-Abelian junction}:
Non-Abelian junctions divide a set of three vacua 
$\left<\ \underline{\cdots}\ AB\right>$,
$\left<\ \underline{\cdots}\ AC\right>$ and
$\left<\ \underline{\cdots}\ BC\right>$
with different labels in two
color components.
Since differences of labels between any pairs of 
these three vacua are in only one color component, 
each constituent wall of the junction is a single wall.
This set of vacua can exist only in the non-Abelian 
gauge theory. 
Therefore we call this type of the 3-pronged 
junction the {\it non-Abelian} junction.
\end{itemize}

\begin{figure}[ht]
\begin{center}
\begin{tabular}{ccc}
\includegraphics[height=4cm]{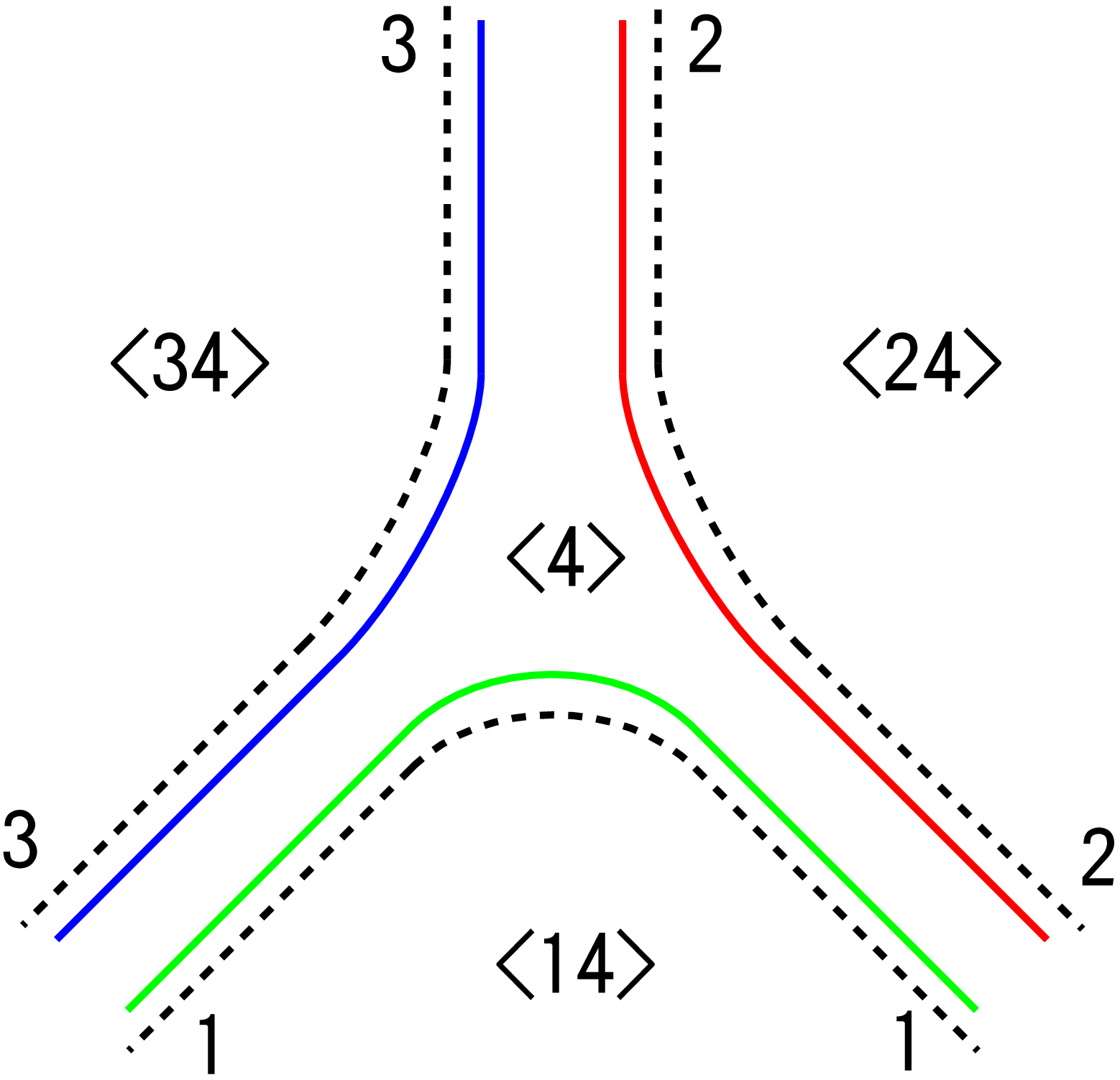}
&\qquad\qquad&
\includegraphics[height=4cm]{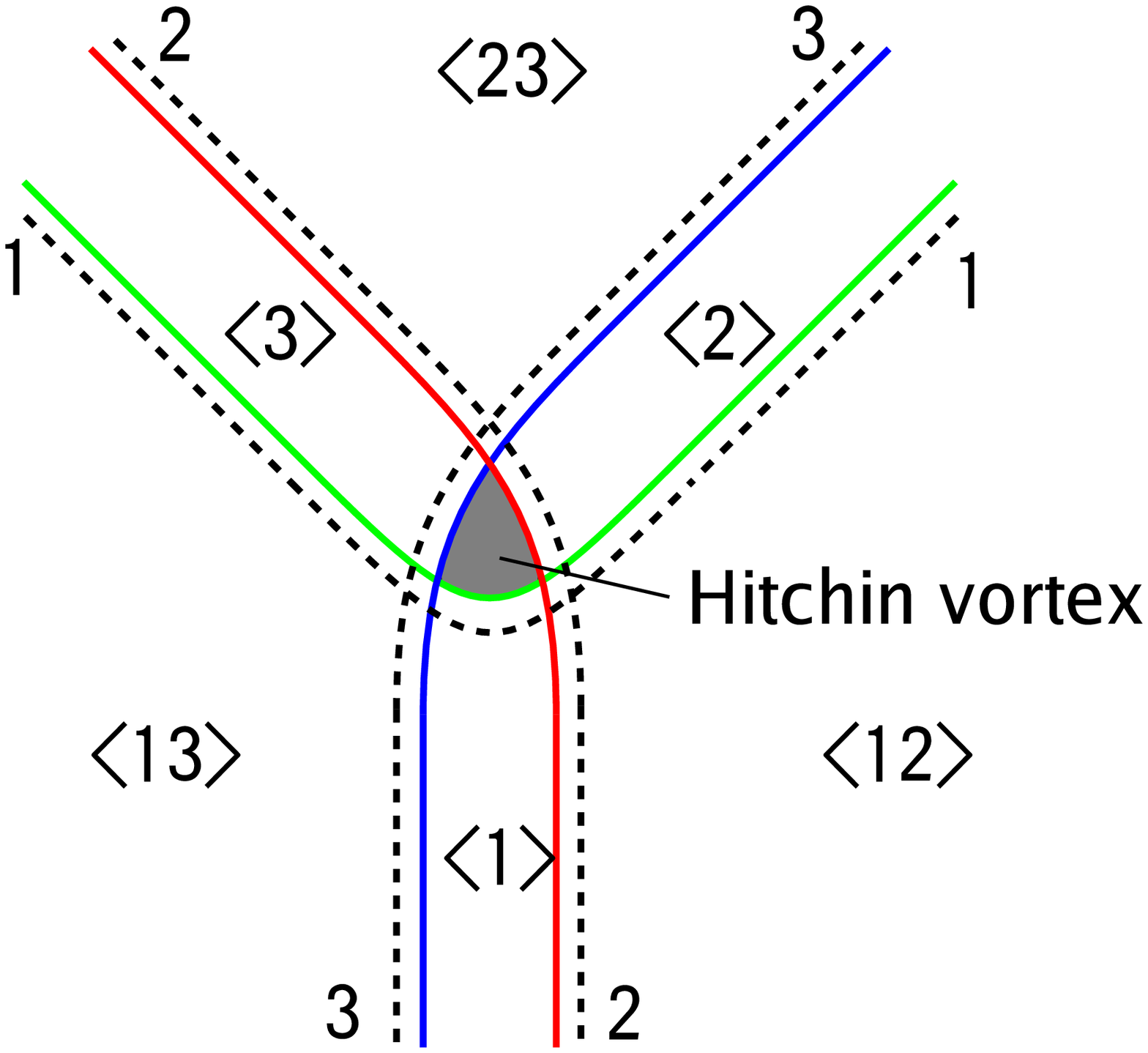}\\
\small{\sf (a) Abelian junction} & & \small{\sf (b) non-Abelian junction}
\vspace*{-.3cm}
\end{tabular}
\caption{\small{\sf Internal structures of the junctions
 with $g \sqrt c \ll |\Delta m+i\Delta n|$}}
\label{hitchin}
\vspace*{-.3cm}
\end{center}
\end{figure}

Let us consider the simplest case of $\NC=2$ and $\NF=4$ to 
explain the difference between the Abelian and the 
non-Abelian wall junctions.
The model has $_4C_2=6$ discrete vacua 
$\langle 12 \rangle$, $\langle 23 \rangle$, 
$\langle 13 \rangle$,
$\langle 14 \rangle$, $\langle 24 \rangle$ and 
$\langle 34 \rangle$. 
The 1/4 BPS wall junction interpolating 
the three vacua $\langle 14 \rangle$, 
$\langle 24 \rangle$ and $\langle 34 \rangle$ is 
an Abelian junction 
while that interpolating 
$\langle 12 \rangle$, $\langle 23 \rangle$ and 
$\langle 13 \rangle$
is a non-Abelian junction.
Internal structures of these junctions are schematically 
shown in Fig.~\ref{hitchin}.
The Abelian junction shown in Fig.~\ref{hitchin}(a) separates
the three different Higgs vacua 
$\langle 14 \rangle$, 
$\langle 24 \rangle$ and $\langle 34 \rangle$.
In the limit with $g\sqrt c \ll |\Delta m+i\Delta n|$, 
each component domain wall
of the junction has internal structure (three-layer structure) 
as explained in Fig.~\ref{wall}(a). 
The same $U(1)$ subgroup is recovered in all 
three middle layers, as denoted by $\langle 4\rangle$. 
They are connected at the junction point so that the 
middle layer of the wall junction is also in the same 
phase $\langle 4 \rangle$ 
as can be seen in Fig.~\ref{hitchin}(a).
The Abelian junction charge $Y$ given in 
Eq.(\ref{Y-charge}) gives always a negative 
contribution to the energy density~\cite{Eto:2005cp}, which 
can be understood as binding energy of the walls. 

On the other hand, the non-Abelian junction has a complicated
internal structure as shown in Fig.~\ref{hitchin}(b).
Although it also separates three different vacua 
$\langle 12 \rangle$, $\langle 23 \rangle$ and $\langle 13 \rangle$,
their middle layers preserve different $U(1)$ subgroups, 
$\langle1\rangle$, $\langle2\rangle$ and $\langle3\rangle$ as 
in Fig.~\ref{hitchin}(b). 
However, all the Higgs fields (hypermultiplets) vanish when 
all three middle layers overlap near the junction point 
so that only the $U(2)$ vector multiplet scalar $\Sigma$ is 
nonvanishing there. 
The $SU(2)$ part of this $\Sigma$ is 
responsible for the ``positive'' $Y$-charge
which might be somewhat 
surprising because 
$Y$-charges 
were negative in 
all the junctions that have been constructed so far 
\cite{Oda:1999az,Eto:2005cp}. 
Since such a positive contribution cannot be interpreted 
as binding energy, we wish to give another understanding. 
The key observation is that the 1/4 BPS equations given 
in Eq.(\ref{bps_eq}) 
include the 1/2 BPS Hitchin equations 
\be
F_{12}=i\left[\Sigma_1,\Sigma_2\right],\quad
\D_1\Sigma_2 - \D_2\Sigma_1 = 0,\quad
\D_1\Sigma_1 + \D_2\Sigma_2 = 0 
\label{hitchin_eq}
\ee
of the Hitchin system, 
if we take 
the traceless part of Eq.(\ref{bps_eq}) and 
ignore the hypermultiplet (Higgs) scalars. 
This reduction occurs at the core of the 
non-Abelian junction, since hypermultiplet scalars 
vanish and hence $U(1)$ and $SU(2)$ part of 
$U(2)$ decouple. 
Therefore the system reduces to the Hitchin system of 
$SU(2)$ subgroup in the middle of the non-Abelian junction. 
Furthermore, the $Y$-charge in Eq.(\ref{Y-charge}) 
completely agrees with the charge the Hitchin system 
 \cite{Eto:2005sw}.
Now we can realize that the positive $Y$-charges of the 
non-Abelian junctions are the charges of  the Hitchin system. 
One might suspect that such solution is not regular because 
the adjoint scalars of Eq.(\ref{hitchin_eq}) grow 
exponentially and their charges diverge~\cite{Cherkis:2000cj}. 
However, as we show below, the $Y$-charge 
for the non-Abelian junctions is finite because the adjoint 
scalar is out of the vacuum value only in a finite
region around the junction point.

\begin{figure}[ht]
\begin{center}
\begin{tabular}{ccc}
\includegraphics[height=3cm]{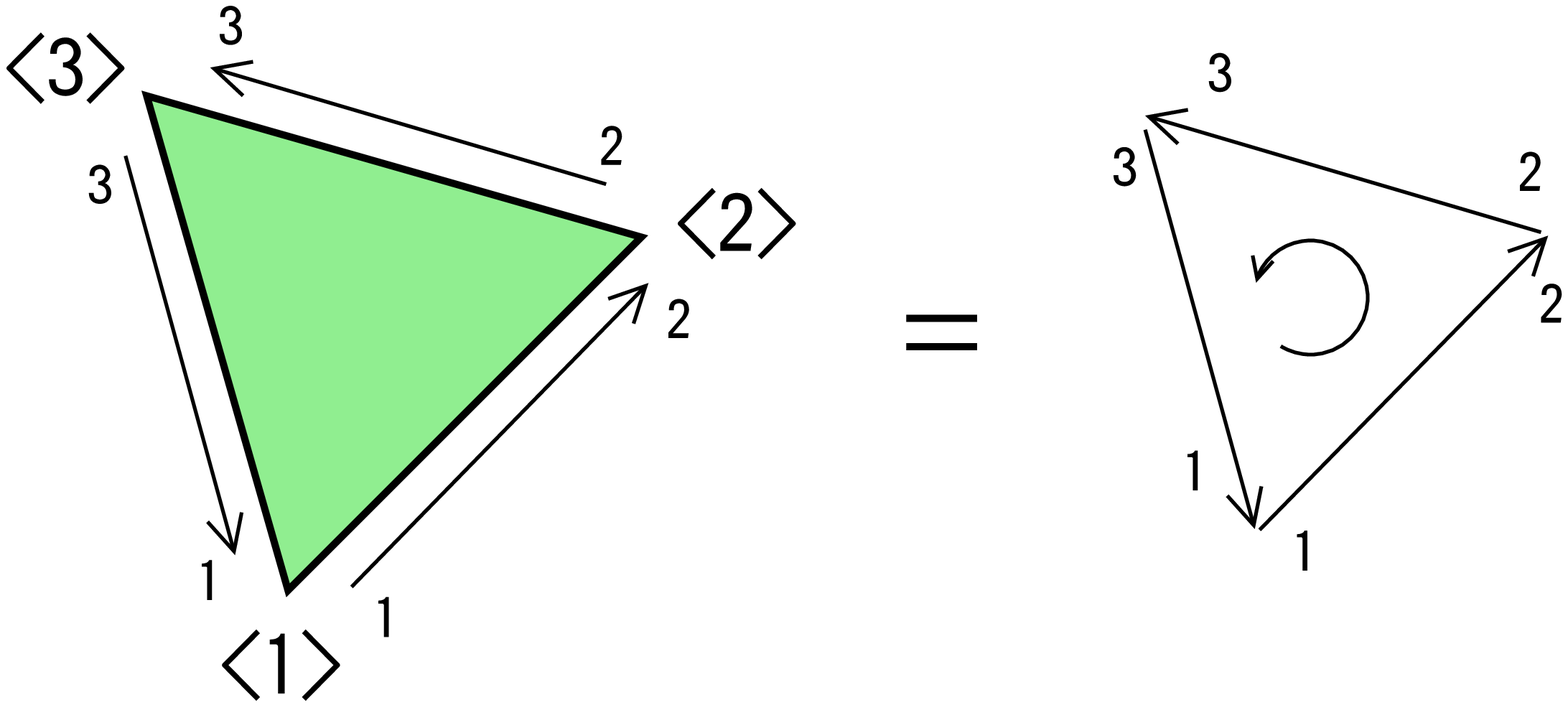}
&\qquad\qquad&
\includegraphics[height=3cm]{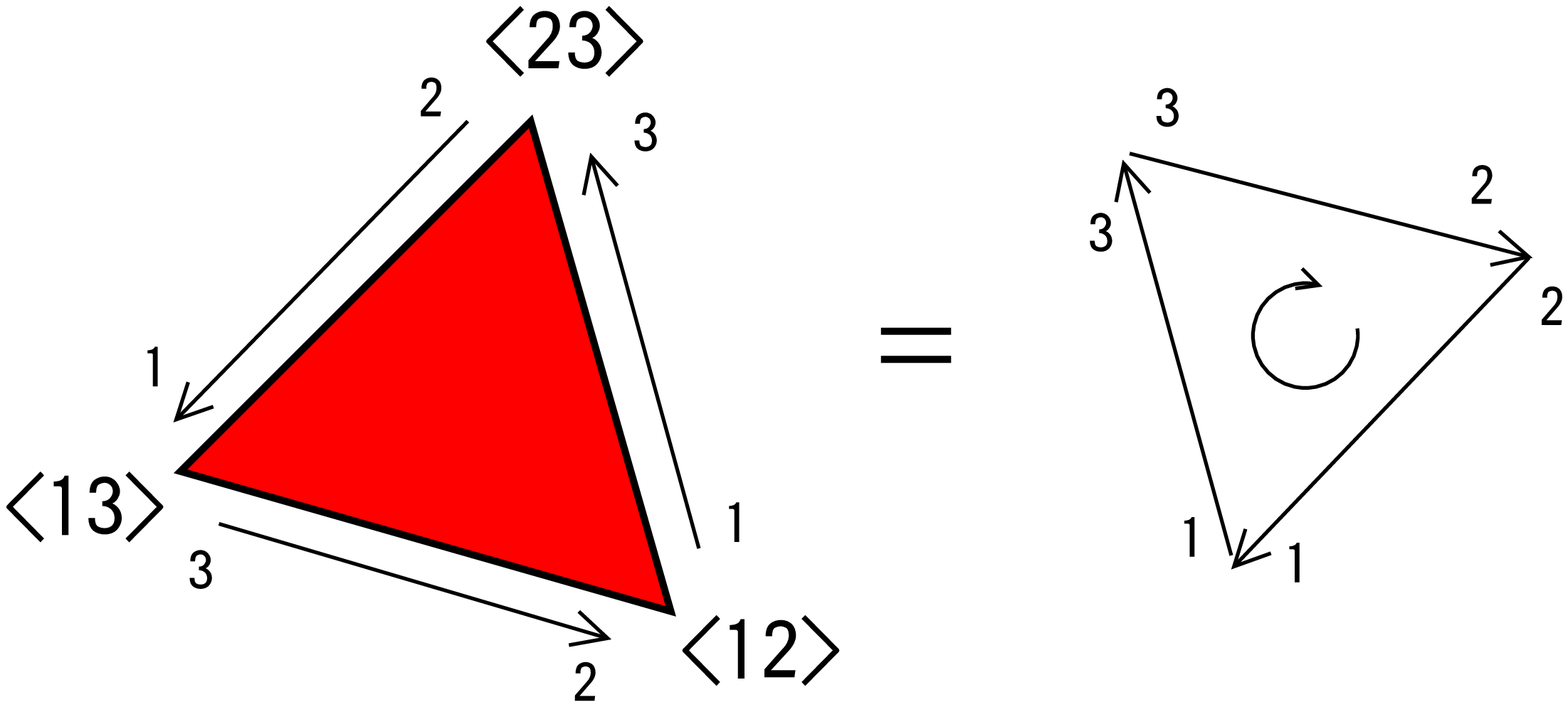}\\
\small{\sf (a) Abelian junction} & & \small{\sf (b) non-Abelian junction}
\vspace*{-.3cm}
\end{tabular}
\caption{\small{\sf grid diagram and their orientation}}
\label{grid}
\end{center}
\vspace*{-.5cm}
\end{figure}

Let us next explicitly show that the Abelian junction has 
negative  $Y$-charge whereas
the non-Abelian junction has positive  $Y$-charge.
We found that domain walls and their junctions in the 
Abelian gauge theory are naturally described 
in the complex $\Sigma$ plane 
\cite{Eto:2005cp}.  
In that plane the SUSY vacuum $\langle A \rangle$ 
is expressed by a point $\mu_A$ and a wall interpolating between
vacua $\langle A \rangle$ and $\langle B\rangle$ is 
by a segment between two points $\mu_A$ and $\mu_B$. 
A diagram made of these points and segments is called 
the grid diagram. 
Furthermore, the 3-pronged junction which divides three 
vacua $\langle A \rangle$, $\langle B\rangle$ and 
$\langle C\rangle$ corresponds to 
a triangle $\triangle ABC$ in that plane.  
In the previous paper \cite{Eto:2005cp} we showed that 
the $Y$-charge of the Abelian junction 
is negative and its magnitude is proportional to the 
area of the triangle. 
Here we again show this fact by a little bit different way. 
To this end, first we rewrite the $Y$-charge in 
Eq.(\ref{Y-charge})
by using the Stokes theorem as
\be
Y = - \frac{2}{g^2}\oint dx^\alpha\ \Tr\left(
\Sigma_1\overset{\leftrightarrow}{\D}_\alpha\Sigma_2\right).
\label{Y-contour}
\ee
We consider the model with $\NF=3$ and
name three vacua $\langle 1\rangle$, $\langle 2\rangle$ 
and $\langle 3\rangle$ {\it counterclockwisely} as in Fig.~\ref{grid}(a).
The 1/4 BPS junction consists of three 1/2 BPS domain walls,
so it can be dealt with as a collection of 
the 1/2 BPS walls in the 
contour integral (\ref{Y-contour}) at three spatial 
infinities.
Therefore the contour integral becomes 
the sum of the three line integrals 
\be
\oint dx^\alpha\ 
\Sigma_1\overset{\leftrightarrow}{\partial}_\alpha\Sigma_2
= \sum_{A=1}^3
T_{\left<A\right>\to \left<A+1\right>},\qquad
T_{\left<A\right>\to \left<A+1\right>}
\equiv
\int_{\left<A\right>}^{\left<A+1\right>}\!\!
dx^\alpha\ \Sigma_1 \overset{\leftrightarrow}
{\partial}_\alpha\Sigma_2,
\label{line_int}
\ee
with identifying $\left<4\right>=\left<1\right>$.
Here the integral $\int_{\left<A\right>}^{\left<A+1\right>}$ 
symbolically expresses
the line integral on the line 
parallel to the vector $\vec n = (m_{A+1}-m_{A},n_{A+1}-n_{A})$.
Although we cannot exactly solve the 1/2 BPS equation
for the finite gauge coupling, 
it is enough to recall that
1/2 BPS solutions are mapped into segments between vacua 
$\mu_A$ and $\mu_{A+1}$
in the complex $\Sigma$ plane. 
Then the 1/2 BPS wall solution interpolating
between vacua $\langle A \rangle$ and 
$\langle A+1 \rangle$ can be 
written as
\be
\Sigma_{\langle A \rangle \to \langle A+1 \rangle}^{U(1)} 
= (\mu_{A+1} - \mu_A)\rho(x^\alpha) + \mu_A,\quad
W_\alpha = 0.
\label{rho}
\ee
with a real function $\rho(x^\alpha)$ tending to 
$\rho(x^\alpha) = 0$ as $x^\alpha \to \langle A\rangle$
and $\rho(x^\alpha) = 1$ as $x^\alpha \to \langle A+1\rangle$.
Plugging this into $T_{\left<A\right>\to \left<A+1\right>}$ 
in Eq.(\ref{line_int}), 
we find $
T_{\left<A\right>\to \left<A+1\right>} 
= \frac{1}{2}\vec{\mu}_A \times \vec{\mu}_{A+1}
$, which is expressed as an exterior product of 2-vectors 
giving a scalar.
Summing $T_{\left<A\right>\to \left<A+1\right>}$ over index $A$,
we can verify that the $Y$-charge is negative 
and proportional to the area of the grid diagram
\be
Y
= -\frac{1}{g^2}\left(\vec\mu_1-\vec\mu_3\right)\times
\left(\vec\mu_2-\vec\mu_3\right)
= -\frac{2}{g^2} \times \left(\text{Area of grid diagram}\right).
\ee

Next let us focus 
on the non-Abelian junction.
We consider $U(2)$ gauge theory with $\NF=3$ flavors whose masses are the same
with those in the Abelian gauge theory.
There exist the three vacua $\langle 12\rangle$, $\langle 23\rangle$
and $\langle 31\rangle$ in this model.
The grid diagrams in the complex $\Sigma$ plane for the Abelian
gauge theory are naturally extended to diagrams in the
$\Tr\Sigma$ plane for the non-Abelian gauge theories.
In the $\Tr\Sigma$ plane the vacua 
$\langle \widetilde A\rangle = \langle BC\rangle$
$(\widetilde A \neq  B \neq C)$ is a point
at $\mu_{\widetilde A} = \mu_B+\mu_C$ and the 1/2 BPS walls correspond to segments
between the two vertices. The non-Abelian junction corresponds
to the triangle $\triangle{\widetilde A}{\widetilde B}{\widetilde C}$ in the plane.
Notice that the triangle $\triangle{\widetilde A}{\widetilde B}{\widetilde C}$
for the non-Abelian junction and the triangle $\triangle ABC$ for the
Abelian junction are congruent  {to each other} 
and coincide when one of them is rotated by the angle $\pi$ 
[see Fig.~\ref{grid}(b)]. 
Furthermore, the orientation for the non-Abelian junction
is opposite to the Abelian junction, namely it is {\it clockwise}.
Similarly to the Abelian junction, we divide
the contour integral in Eq.(\ref{Y-contour}) into
three line integrals as
\be
\oint dx^\alpha\ \Tr\left(
\Sigma_1\overset{\leftrightarrow}{\D}_\alpha\Sigma_2\right)
= \sum_{\widetilde A =1}^3
T_{\langle\widetilde A\rangle\to \langle\widetilde{A+1}\rangle},
\quad
T_{\langle\widetilde A\rangle\to \langle\widetilde{A+1}\rangle} 
\equiv
\int^{\langle\widetilde{A+1}\rangle}_{\langle\widetilde A\rangle}
\!\!
dx^\alpha\ \Tr\left[\Sigma_1\overset{\leftrightarrow}
{\D}_\alpha\Sigma_2\right],
\label{NA_junc}
\ee
with identifying 
$\langle\widetilde 4\rangle=\langle\widetilde 1\rangle$.
In the each line integral the junction can be regarded as 
a 1/2 BPS single wall. 
The single wall interpolating between 
$\langle AB\rangle$ 
and $\langle BC\rangle$ can be obtained by  
embedding the solution of
a single wall interpolating between $\langle A\rangle$
and $\langle C\rangle$ in the Abelian model 
\cite{Isozumi:2004jc} as\footnote{
We need to insert a gauge transformation to connect all 
three wall solutions to form the junction. } 
\be
\Sigma_{\langle AB\rangle \to \langle BC\rangle}
= \left(
\begin{array}{cc}
\Sigma^{U(1)}_{\langle A \rangle \to \langle C\rangle} &\\
& \mu_B
\end{array}
\right),\quad
W_\alpha = 0,
\ee
where $\Sigma^{U(1)}_{\langle A \rangle \to \langle C\rangle}$
is the solution of the single wall given in Eq.(\ref{rho}).
Substituting this solution into Eq.(\ref{NA_junc}), we find that
$T_{\langle\widetilde A\rangle\to \langle\widetilde{A+1}\rangle}$
has a sign opposite to 
$T_{\langle A\rangle\to \langle{A+1}\rangle}$, 
because of the opposite orientation:
$
T_{\langle\widetilde A\rangle\to \langle\widetilde{A+1}\rangle}
= T_{\langle A+1 \rangle\to \langle A\rangle}
= - T_{\langle A\rangle\to \langle{A+1}\rangle}$.
Therefore we conclude that the $Y$-charge for the 
non-Abelian junction
is positive and its magnitude agrees with the area of the 
grid diagram
\be
Y = - \frac{2}{g^2}\sum_{\widetilde A = 1}^{3}
T_{\langle\widetilde A\rangle\to \langle\widetilde{A+1}\rangle}
= \frac{2}{g^2}\times\left(\text{Area of grid diagram}\right).
\ee

\begin{figure}[ht]
\begin{center}
\begin{tabular}{ccc}
\includegraphics[height=3cm]{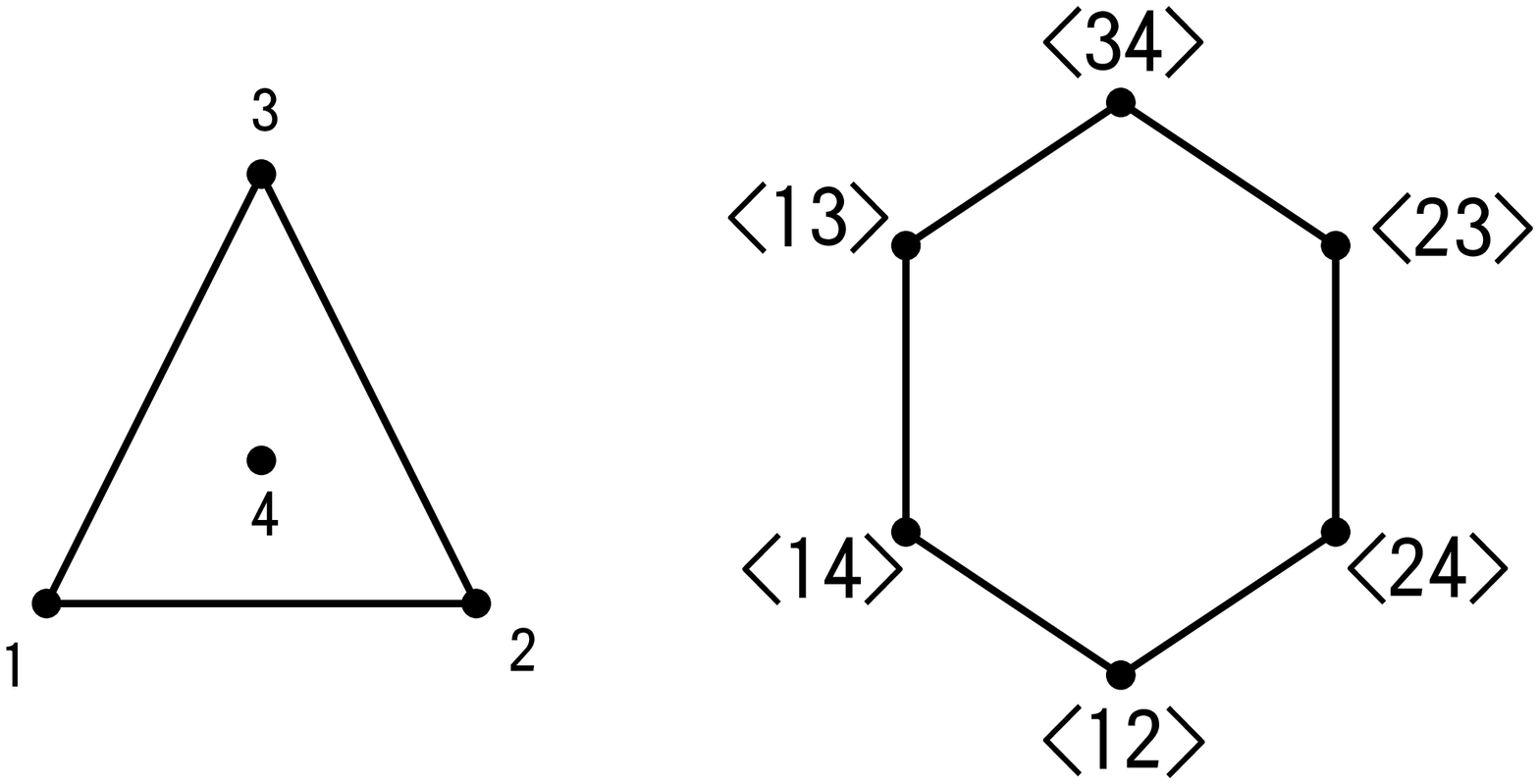}
&\qquad&
\includegraphics[height=2.8cm]{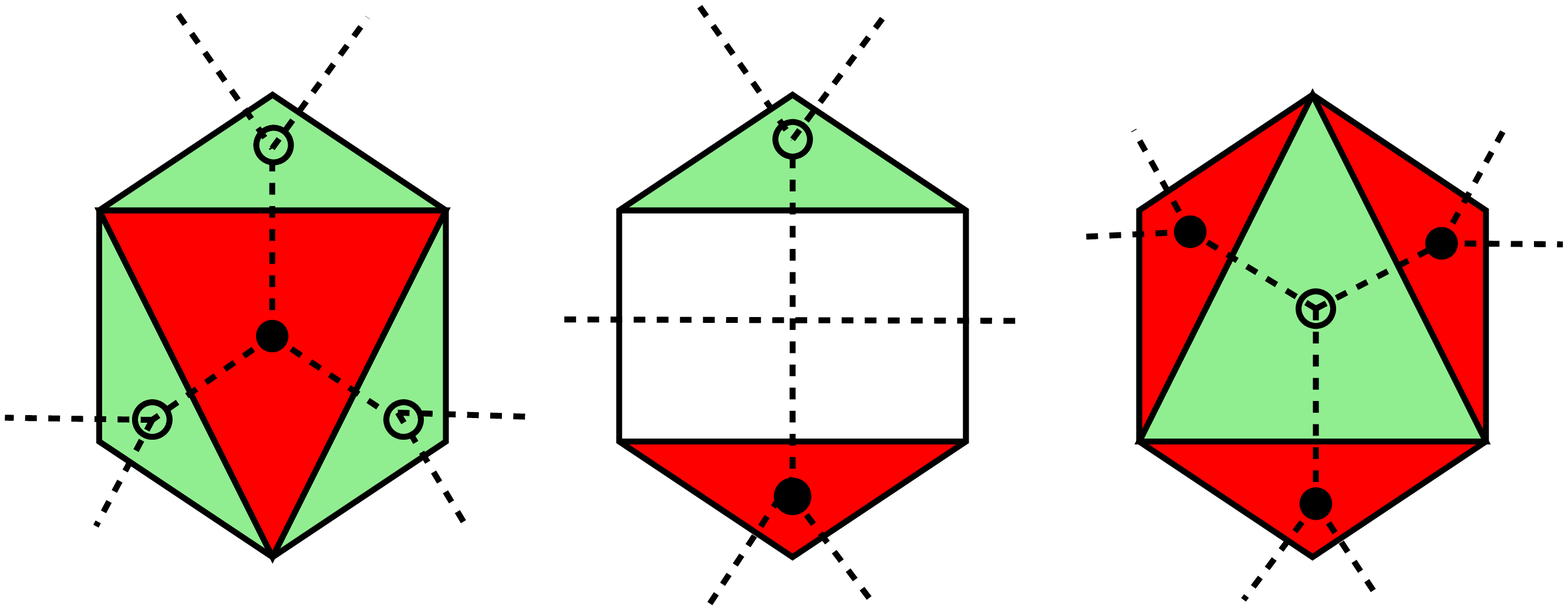}\\
\small{\sf (a1) mass arrangement and vacua} & & 
\small{\sf (a2) grid diagrams and web configurations}\\
\includegraphics[height=2.5cm]{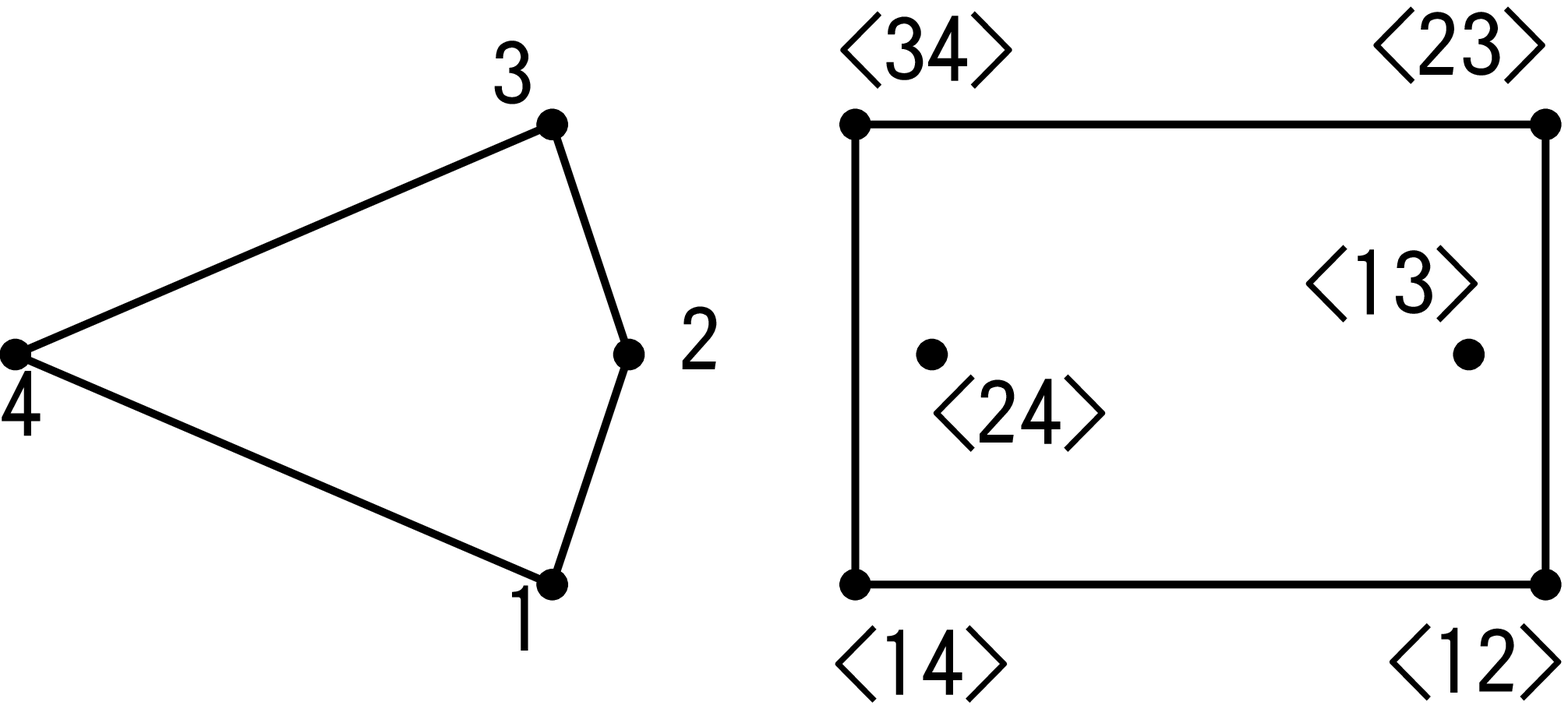}
&\qquad&
\includegraphics[height=3.5cm]{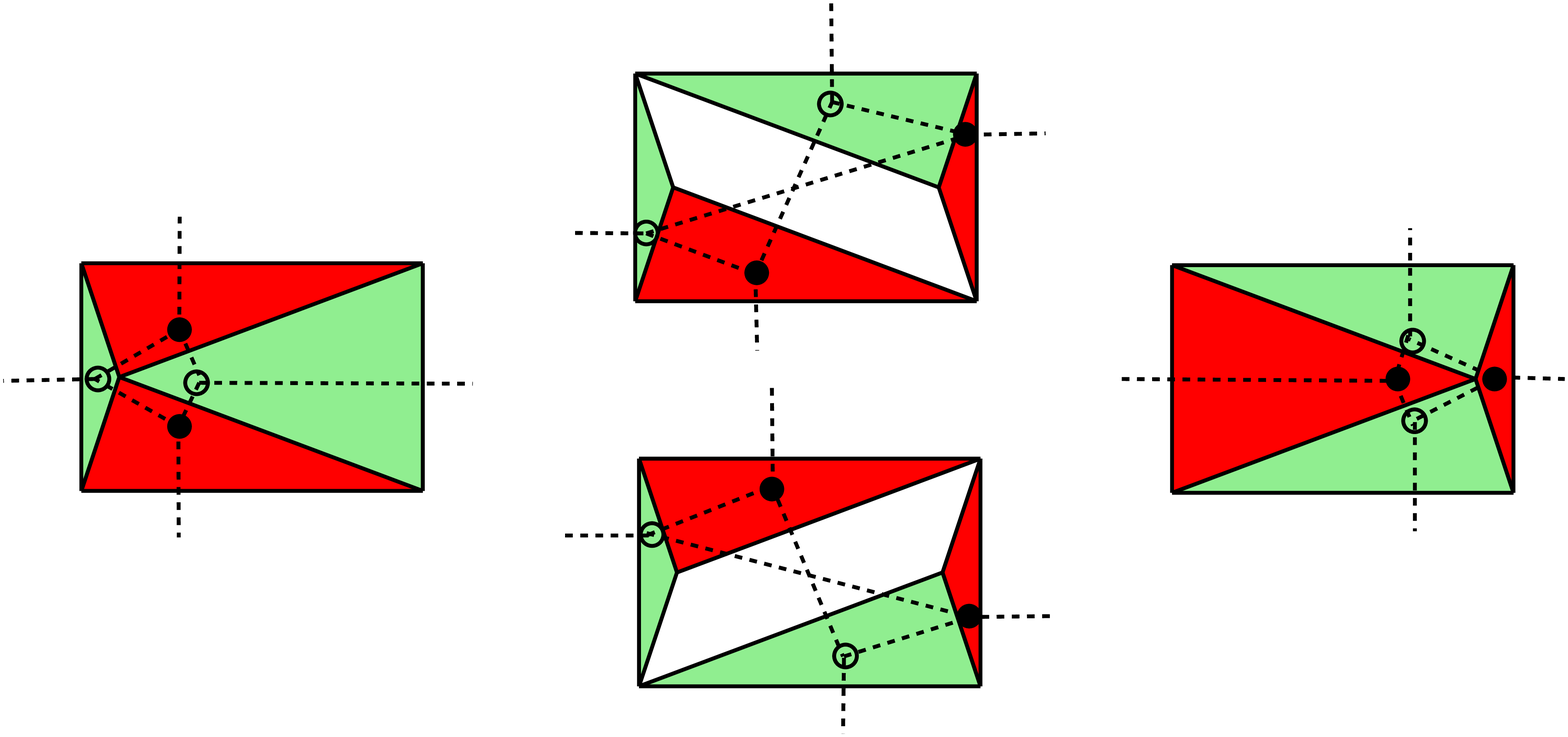}\\
\small{\sf (b1) mass arrangement and vacua} & & 
\small{\sf (b2) grid diagrams and web configurations}
\vspace*{-.3cm}
\end{tabular}
\caption{\small{\sf (a1) and (a2) for the hexagon-type and
(b1) and (b2) for the parallelogram-type web.}}
\label{model}
\vspace*{-.3cm}
\end{center}
\end{figure}

For general $\NF$ and $\NC(<\NF)$ the webs of walls are 
constructed by the Abelian 3-pronged junction, 
the non-Abelian 3-pronged junction, and the 
intersection of penetrable walls 
(with vanishing $Y$-charge) as their 
building blocks. 
The rules for the grid diagrams in the $U(\NC)$ gauge
theory are given as follows: 
\begin{enumerate}[i)]

\item \label{rule1}
Determine mass arrangement $\mu_A$
and plot $_{\NF}C_{\NC}$ vacuum points 
$\langle A_r\rangle$ at 
$\sum_{r=1}^{\NC}\mu_{A_r}$ in the complex $\Tr\Sigma$ plane.

\item \label{rule2}
Draw a convex polygon by choosing a set of vacuum points, which 
determines the 
boundary condition of a BPS solution.
Here each edge of the convex polygon must be a 1/2 BPS single wall
between pairs of the vacuum points
$\langle\ \underline{\cdots}\ A\rangle$ and 
$\langle\ \underline{\cdots}\ B\rangle$.

\item \label{rule3}
Draw all possible internal segments within the convex polygon
describing 1/2 BPS single walls
forbidding any segments to cross.

\item \label{rule4}
 Identify Abelian triangles with vertices  
$\langle\ \underline{\cdots}\ A\rangle$, $\langle\ 
\underline{\cdots}\ B\rangle$
and $\langle\ \underline{\cdots}\ C\rangle$ to
Abelian 3-pronged junctions.
Identify non-Abelian triangles with vertices  
$\langle\ \underline{\cdots}\ AB\rangle$, $\langle\ 
\underline{\cdots}\ BC\rangle$
and $\langle\ \underline{\cdots}\ CA\rangle$ 
to non-Abelian 3-pronged junctions.
 Identify parallelograms with vertices
$\langle\ \underline{\cdots}\ AB\rangle$, 
$\langle\ \underline{\cdots}\ BC\rangle$,
$\langle\ \underline{\cdots}\ CD\rangle$ and 
$\langle\ \underline{\cdots}\ DA\rangle$
to intersections of two penetrable walls with vanishing 
$Y$-charges.

\end{enumerate}
Each grid diagram determines the topology of the web 
configuration and represents a subspace of the moduli space.
The moduli of the 
web configurations are finally specified by 
drawing dual diagrams of the grid diagram, as illustrated 
by an example in Fig.\ref{model}(a2) where the dashed line 
represents a dual diagram.
Configurations with different topologies can be obtained by
changing structure of internal segments and/or choosing other
convex polygons. Thus the total moduli space of the web is 
obtained by gathering all possible grid diagrams.

The $U(2)$ gauge theory with four flavors gives the 
simplest example of wall webs containing 
both Abelian and non-Abelian junctions. 
According to mass arrangements for the hypermultiplets,
the webs can be classified to two classes.
Let us first consider a mass arrangement 
such as in the left figure of Fig.~\ref{model}(a1). 
According to the rule \ref{rule1}), 
vacuum points can be plotted in 
the $\Tr\Sigma$ plane as in the right figure 
of Fig.~\ref{model}(a1). 
Moreover, we connect 
pairs of vertices corresponding
to single walls under the rule  \ref{rule3}), then
we obtain the grid diagram for the web of walls.
In Fig.~\ref{model}(a2) we show three examples of the 
grid diagrams.
Here light shaded (green) triangles are Abelian junctions 
and dark shaded (red) triangles are non-Abelian 
junctions %(rule  \ref{rule4})) 
while 
white parallelograms are intersections of penetrable 
walls (rule  \ref{rule4})).
The dual diagrams in the configuration spaces are 
shown as the broken lines
in Fig.~\ref{model}(a2). These diagrams are transformed 
to each other by changing the moduli parameters of the 
solution as we show in the next section.
The other class of the web configurations is obtained  by 
the mass arrangement given in the left of 
Fig.~\ref{model}(b1).
According to the rule  \ref{rule1}), the vacuum points 
are plotted in the $\Tr\Sigma$ plane 
as in the right figure of Fig.~\ref{model}(b1). 
By connecting the line between 
several pairs under the rule  \ref{rule3}) and painting 
triangles two colors for Abelian 
and non-Abelian junctions, we get the grid diagrams. 
We show several examples in Fig.~\ref{model}(b2). 
Unlike the above case, the grid diagram has 4 edges, so
the web diagrams (shown by broken lines) has 4 external legs.
These webs have a loop therein.

\begin{figure}[ht]
\begin{center}
\begin{tabular}{ccccc}
\includegraphics[height=3.3cm]{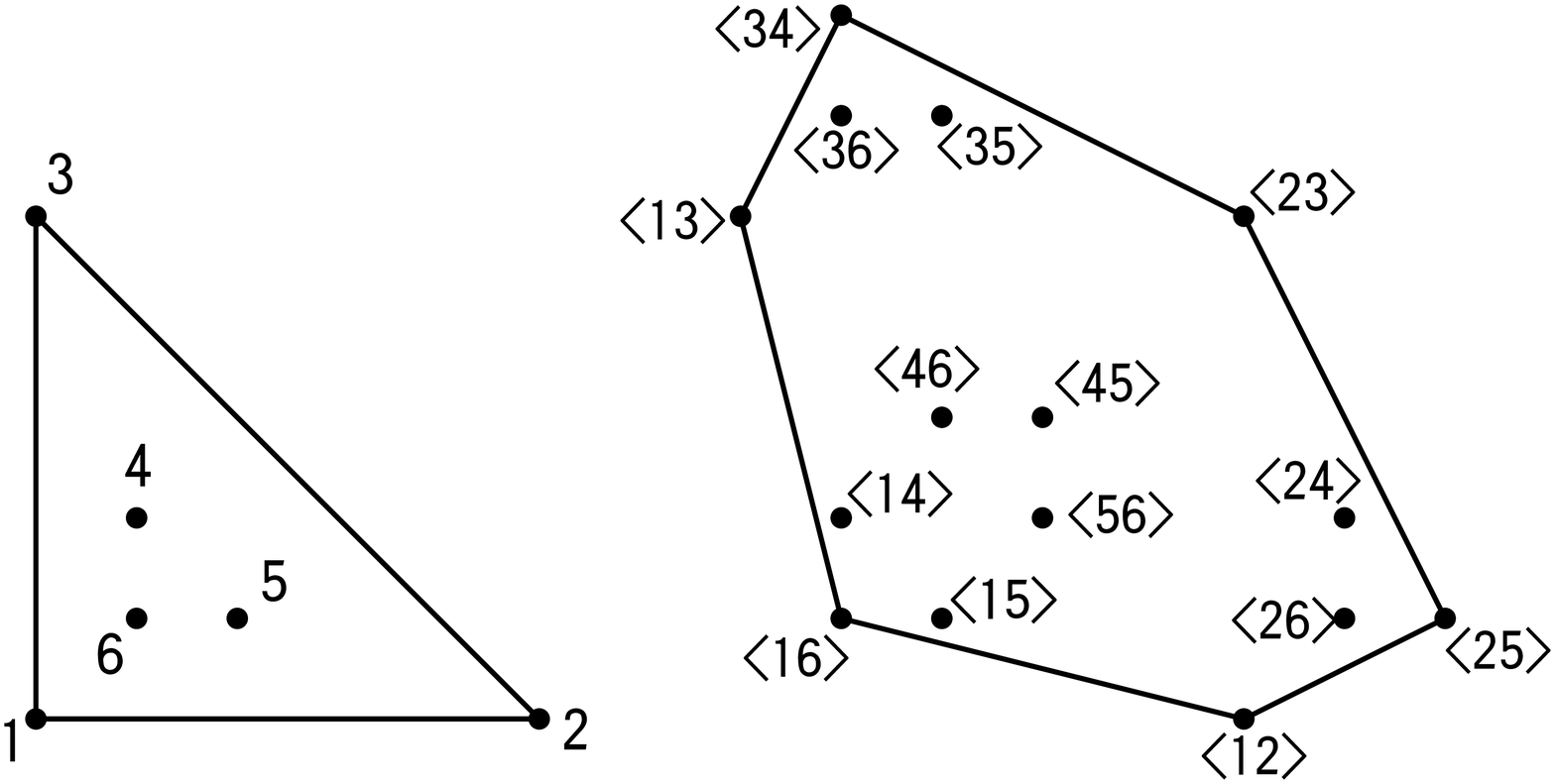} &\qquad&
\includegraphics[height=3.3cm]{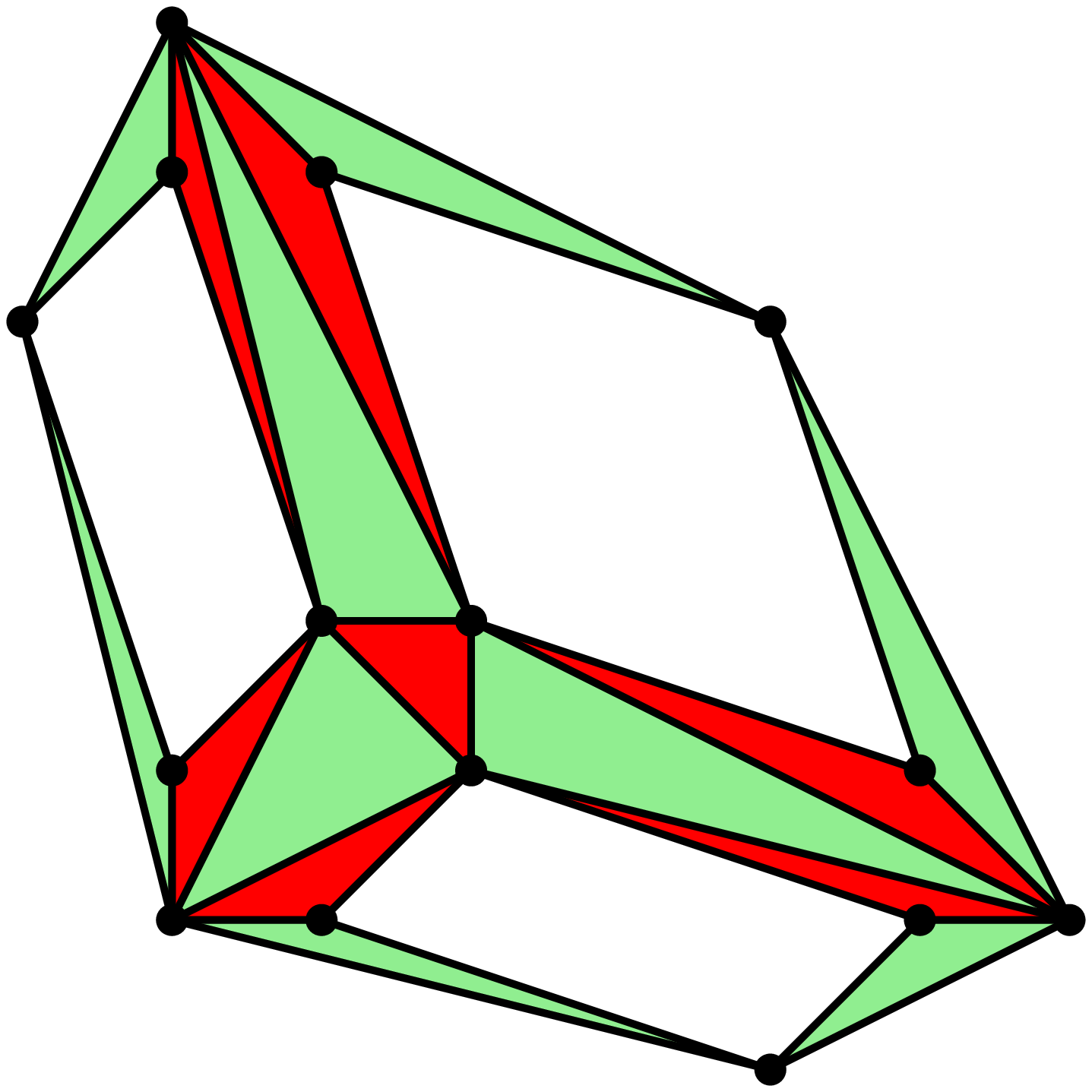} &\qquad&
\includegraphics[height=3.8cm]{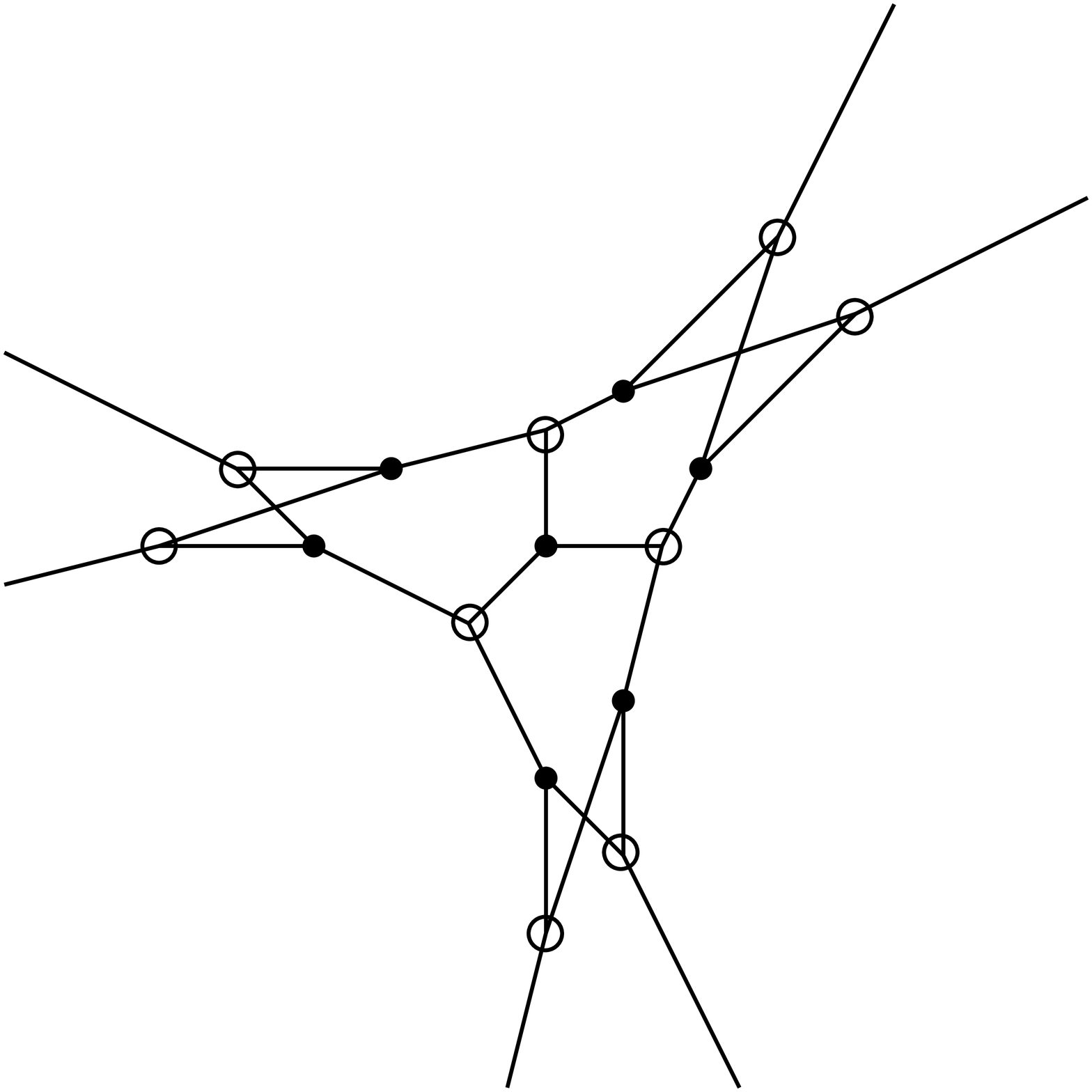} \\
\small{\sf (a)  mass arrangement and vacua} &&
\small{\sf (b)  grid diagram} && \small{\sf (c) web diagram}
\vspace*{-.3cm}
\end{tabular}
\caption{\small{\sf The web of walls in the $U(2)$ gauge theory with 6 flavors.}}
\label{general}
\end{center}
\vspace*{-.5cm}
\end{figure}

More complicated webs can easily be understood in terms
of the grid diagrams in the complex $\Tr\Sigma$ plane. 
A web in the model of $(\NF,\NC)=(6,2)$ with $_6C_2=15$ vacua 
is shown in Fig.~\ref{general}.
The grid diagram has 6 edges, 9 Abelian triangles, 
7 non-Abelian triangles
and 3 parallelograms.

%%%%%%%%%%%%%%%%%%%%%%%%%%%%%%%%%%%%%%%%%%%%%%%%%%%%%%%%%%%%%%%%
\subsubsection*{Moduli and Configurations}

Generic configurations of the 1/4 BPS webs are made of 
several Abelian and non-Abelian junctions.
In order to investigate such generic webs more closely, 
we shall clarify the structure 
of the moduli parameters in generic solutions 
to Eq.(\ref{bps_eq}).
All the solutions of the 1/4 BPS equations can be expressed 
by matrices 
$S(x^1,x^2)$ and $H_0$, which is called 
the moduli matrix. 
The invertible $\NC\times\NC$ matrix $S$ is a 
function of $x^1$ and $x^2$, and $H_0$ 
is an $\NC\times\NF$ complex constant matrix 
with rank $\NC$.  
Generic solutions to the first two equations in 
Eq.~(\ref{bps_eq}) are given in \cite{Eto:2005cp} by 
\be
 H = S^{-1}H_0e^{M_1x^1+M_2x^2},\quad
 W_\alpha - i\Sigma_\alpha= -iS^{-1}\partial_\alpha S,\quad
 \left(\alpha = 1,2\right).
 \label{sol}
\ee
The last equation of Eq.(\ref{bps_eq}) can be rewritten into 
a gauge invariant equation, 
which is called the master equation \cite{Eto:2005cp} 
and determines 
$S$ for a given $H_0$ up to a gauge 
transformation. 
All the elements of the moduli matrix $H_0$ are 
integration constants, and can be moduli parameters 
in the solutions. 
However $(H_0,S) \sim V(H_0,S)$ with $V \in GL(\NC,{\bf C})$
gives the same physical fields in Eq.(\ref{sol}), 
and defines an equivalence relation which we 
call $V$-equivalence relation in what follows. 
Therefore we obtain the moduli space of 
the 1/4 BPS equations (\ref{bps_eq}) to be the 
complex Grassmann manifold  
$
{\cal M}_{\rm tot} \simeq
G_{\NF,\NC} = \{H_0\ |\ H_0 \sim V H_0,\ V\in GL(\NC,{\bf C})\}
$. 
We call ${\cal M}_{\rm tot}$ the {\it total moduli space} 
because it contains all possible topological sectors 
of 1/2 and 1/4 BPS states and vacua (1/1 BPS) \cite{Eto:2005cp}.

When a point in the moduli space $G_{\NF,\NC}$ is given,
we can figure out the configuration of the 
corresponding 1/4 BPS web as follows.
The energy density of the web is 
approximately given in terms of 
the moduli matrix $H_0$ as
\be
{\cal E} \simeq \frac{c}{2}
\left(\partial_1^2+\partial_2^2\right)
\log \det 
\left(H_0 e^{2(M_1x^1+M_2x^2)} H_0^\dagger\right).
\ee
Notice that this expression is strictly correct for the 
strong coupling limit where we can exactly solve the 
1/4 BPS equations (\ref{bps_eq}) \cite{Eto:2005cp}.
We can rewrite this expression in a more useful form 
\be
{\cal E} = \frac{c}{2}
\left(\partial_1^2+\partial_2^2\right)
\log 
\sum_{\langle A_r\rangle}
e^{2{\cal W}^{\langle A_r\rangle}},
\label{E_estimate}
\ee
where we define a linear function 
${\cal W}^{\langle A_r\rangle}$ 
for each vacuum $\langle A_r\rangle$ as
\be
{\cal W}^{\langle A_r\rangle}(x^1,x^2)
\equiv
\sum_{r=1}^{\NC} \left(m_{A_r}x^1 + n_{A_r}x^2\right)
+ a^{\langle A_r\rangle}.
\ee
Here $e^{a^{\langle A_r\rangle}}$ gives
the real part of the Pl\"ucker coordinates\footnote{
The $\NC(\NF-\NC)$ dimensional complex 
Grassmann manifold $G_{\NF,\NC}$ can be embedded into
the $_{\NF}C_{\NC} - 1$ 
dimensional complex projective space 
${\bf C}P^{_{\NF}C_{\NC}-1}$ by the Pl\"ucker embedding.
The coordinates $\det H_0^{\langle A_r\rangle}$ 
of ${\bf C}P^{_{\NF}C_{\NC}-1}$
are called the Pl\"ucker coordinates which satisfies
an equivalence relation 
$
(\cdots,\det H_0^{\langle A_r\rangle},\cdots)
\sim \det V (\cdots,\det H_0^{\langle A_r\rangle},\cdots)
$ (see the footnote \ref{plucker}).
} 
of $G_{\NF,\NC}$, given by 
$\det H_0^{\langle A_r\rangle}$, 
where $H_0^{\langle A_r\rangle}$ 
is $\NC\times\NC$ matrix whose elements are given by
$(H_0^{\langle A_r\rangle})^s{_t}= (H_0)^s{_{A_t}}$:
\be
a^{\langle A_r\rangle}+ib^{\langle A_r\rangle}
\equiv \log \det H_0^{\langle A_r\rangle}.
\ee
We call $e^{{\cal W}^{\langle A_r\rangle}}$ in the logarithm 
in Eq.(\ref{E_estimate}) the weight of the vacuum 
$\langle A_r\rangle$.
For the vacuum configuration $\langle A_r\rangle$, 
all the weights of vacua except
for the vacuum $\langle A_r\rangle$ vanishes.
Therefore we find
$\log\left(\sum e^{2{\cal W}}\right) 
= 2{\cal W}^{\langle A_r\rangle}$
and then
the energy density, of course, vanishes.
When there are several domains of the vacua,
domain walls appear as transition lines 
between the domains.
At the vacuum domain $\langle A_r\rangle$
the weight $e^{{\cal W}^{\langle A_r\rangle}}$ is dominant
compared to other weights.
There $\log\left(\sum e^{2{\cal W}}\right)$
in Eq.(\ref{E_estimate}) is an almost linear function 
$
\log\left(\sum e^{2{\cal W}^{\langle A_r\rangle}}\right) \sim 
\max\left[\cdots, 2{\cal W}^{\langle A_r\rangle},\cdots\right],
$
so the energy density vanishes there.
Only around transition lines between different domains,
the energy density can have nonzero values. 
Thus positions of the single walls can be estimated
by the condition of  {equating} weights of the vacua.
For example, the position of a single wall 
separating the two vacua $\langle \ \underline{\cdots}\ A \rangle$
and $\langle \ \underline{\cdots}\ B \rangle$ is given by
the condition ${\cal W}^{\langle \underline{\cdots} A\rangle}
= {\cal W}^{\langle \underline{\cdots} B\rangle}$:
\be
(m_A-m_B)x^1 + (n_A - n_B)x^2 
+ a^{\langle \underline{\cdots} A\rangle}
- a^{\langle \underline{\cdots} B\rangle} = 0.
\label{wall_line}
\ee
The position of a 3-pronged junction can be 
estimated as an intersecting point of 
three lines (\ref{wall_line}) for 
the three constituent single walls of that junction. 
In other word,
it can be estimated by 
equating three weights of vacua, like 
${\cal W}^{\langle \underline{\cdots} A\rangle}
= {\cal W}^{\langle \underline{\cdots} B\rangle}
= {\cal W}^{\langle \underline{\cdots} C\rangle}
$ for Abelian junctions 
or 
$
{\cal W}^{\langle \underline{\cdots} AB\rangle}
= {\cal W}^{\langle \underline{\cdots} BC\rangle}
= {\cal W}^{\langle \underline{\cdots} AC\rangle}
$ for non-Abelian junctions.
Eq.(\ref{wall_line}) means that slopes of single walls
are determined by mass differences between adjacent vacua, 
which agrees with the previous result of the central charge 
in Eq.(\ref{central_charge}).
Besides slopes of walls,
Eq.(\ref{wall_line}) contains additional data about 
positions of the single walls described by moduli 
$a^{\langle A_r\rangle}$.
Thus we can attribute the deformation of the shape of the 
web to the change of the moduli parameters. 
To do this, however, one has to note that 
all the parameters $a^{\langle A_r\rangle}$ 
are not independent since the Pl\"ucker coordinates 
$\det H_0^{\langle A_r\rangle}$ must satisfy 
$2(_{\NF}C_{\NC}-1-\NC(\NF-\NC))$ identities 
called the Pl\"ucker relations.\footnote{
\label{plucker}
The Pl\"ucker relations for embedding $G_{\NF,\NC}$ into
${\bf C}P^{_{\NF}C_{\NC}-1}$ are known as 
$\sum_{k=0}^{\NC} (-1)^k
\det\left( H_0^{\langle A_1\cdots A_{\NC-1}B_k\rangle} \right)
\det\left( H_0^{\langle B_0\cdots 
\check{B}_k\cdots B_{\NC}\rangle} \right)
=0$ where the check over $B_k$ denotes removing $B_k$ from 
$\langle B_0\cdots B_k\cdots B_{\NC}\rangle$. 
However 
only $2(_{\NF}C_{\NC} - 1 - \NC (\NF-\NC))$ 
of them are independent.
}

{\it Examples:}
As a concrete example we examine the wall webs 
in the $\NF=4,\NC=2$ model in the rest of this paper. 
The $G_{4,2}$ can be embedded into ${\bf C}P^5$ with 
one Pl\"ucker relation and can be parameterized by six complex 
parameters $a^{\langle AB\rangle} + ib^{\langle AB\rangle}$ 
of homogeneous coordinates for ${\bf C}P^5$.
Due to the $V$-equivalence relation and the Pl\"ucker 
relation, only four complex moduli parameters are independent. 
Notice that three out of eight real parameters
are Nambu-Goldstone (NG) modes associated with 
the broken global $U(1)^3_{\rm F}$ symmetry. 
The remaining five parameters can change the shape of 
the web, but these three parameters do not.

\begin{figure}[ht]
\begin{center}
\begin{tabular}{ccc}
\includegraphics[height=3.5cm]{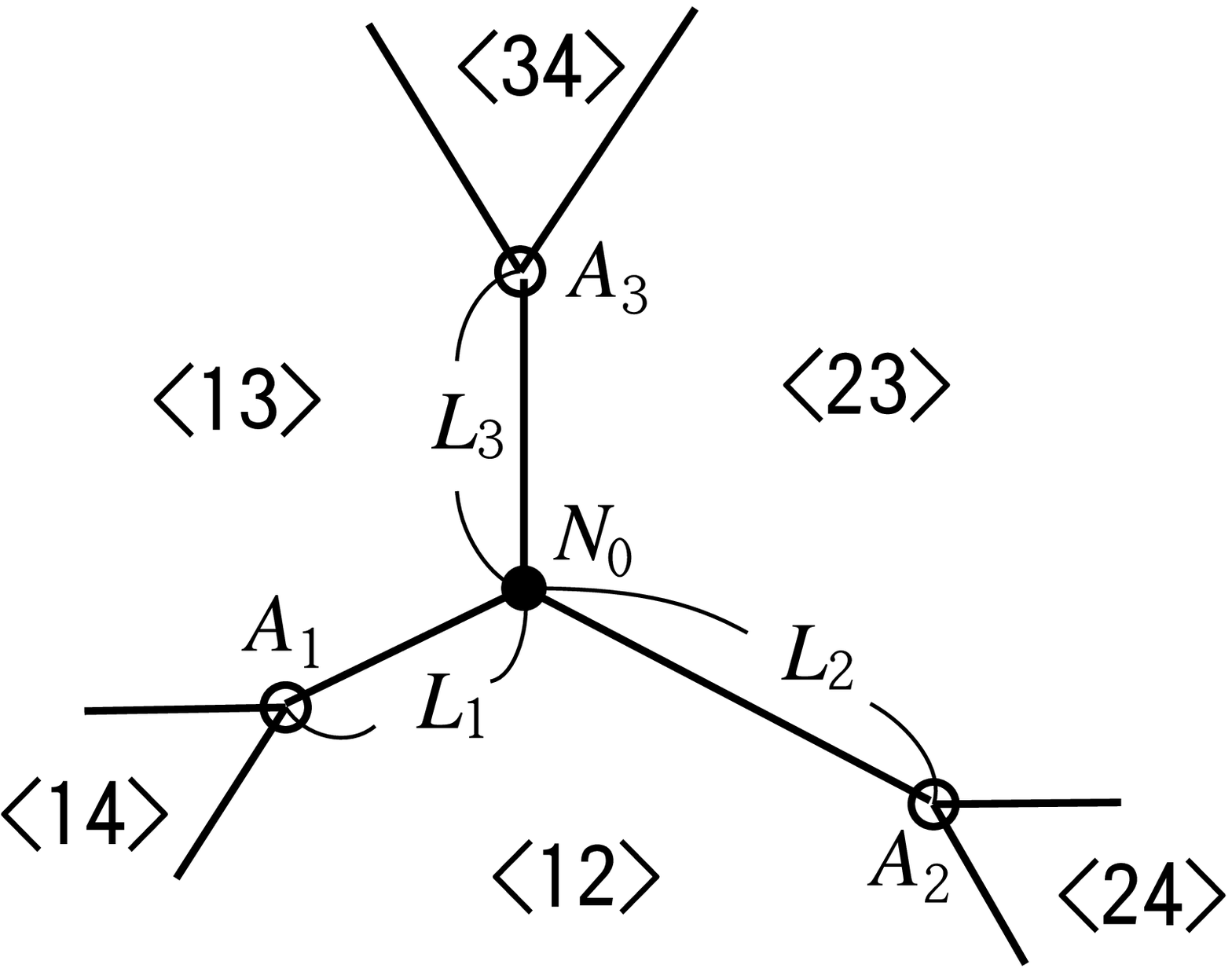}
&\qquad&
\includegraphics[height=3.5cm]{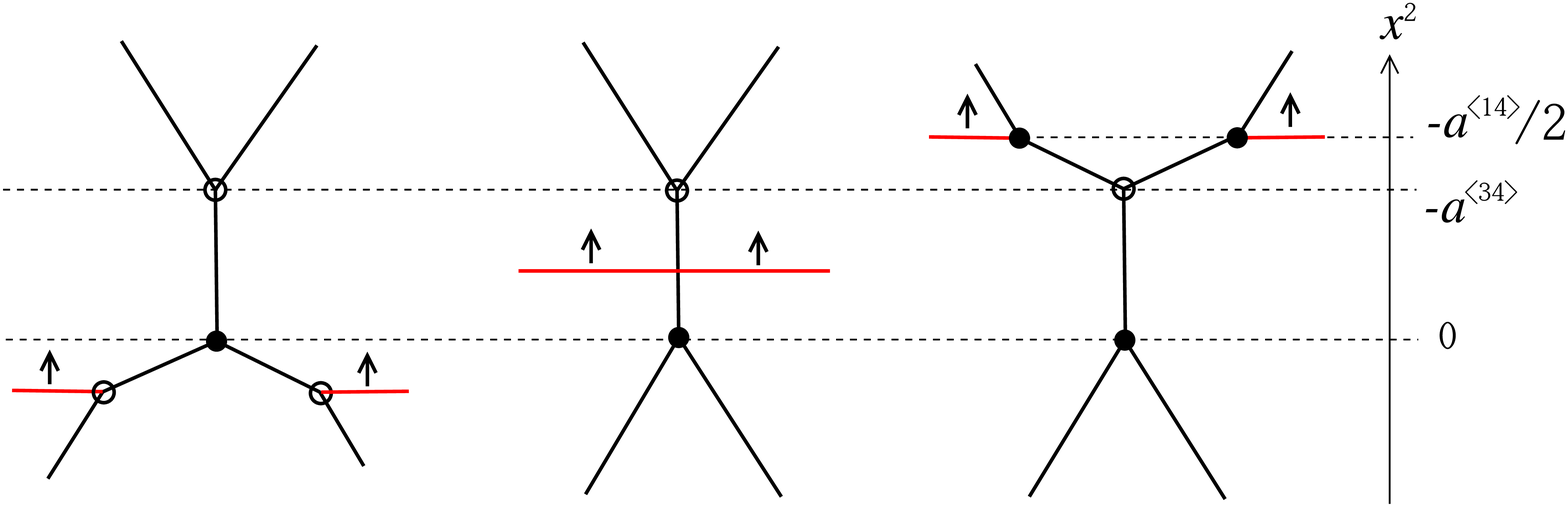}\\
\small{\sf (a) A web configuration} & & 
\small{\sf (b) Changing moduli of the web}
\vspace*{-.3cm}
\end{tabular}
\caption{\small{\sf The web of walls for the 
hexagon-type mass arrangement.}}
\label{hex_mod}
\vspace*{-.3cm}
\end{center}
\end{figure}

\begin{figure}[ht]
\begin{center}
\includegraphics[height=2.5cm]{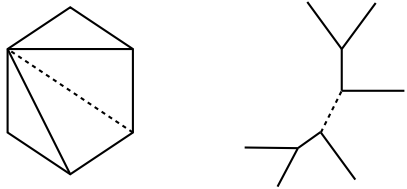}
\caption{\small{\sf A web configuration in ${\bf C}P^5$ model. 
This is forbidden for $G_{4,2}$ model.}}
\label{cp5}
\vspace*{-.3cm}
\end{center}
\end{figure}

First, let us examine the hexagon-type web of the walls with
the mass parameters $M = {\rm diag}(-\sqrt3-i,\sqrt3-i,2i,0)$
like webs given in Fig.~\ref{model}(a), 
followed by the other type in Fig.~\ref{model}(b).
We start with the configuration given in Fig.~\ref{hex_mod}(a)
and denote the complex positions of 
the three outer Abelian junctions by $A_1,A_2,A_3$ and 
that of the center non-Abelian junction by $N_0$.
The positions of the Abelian junctions are 
$A_1=\left(
\frac{-a^{\langle13\rangle}+3a^{\langle14\rangle}-2a^{\langle12\rangle}}{2\sqrt3},
\frac{-a^{\langle13\rangle}+a^{\langle14\rangle}}{2}
\right)$,
$A_2=\left(
\frac{a^{\langle23\rangle}-3a^{\langle24\rangle}+2a^{\langle12\rangle}}{2\sqrt3},
\frac{-a^{\langle23\rangle}+a^{\langle24\rangle}}{2}
\right)$,
$A_3=\left(
\frac{-a^{\langle23\rangle}+a^{\langle13\rangle}}{2\sqrt3},
\frac{a^{\langle23\rangle}+a^{\langle13\rangle}
-2a^{\langle34\rangle}}{2}
\right)$ and that of the non-Abelian junctions is
$N_0=\left(
\frac{-a^{\langle23\rangle}+a^{\langle13\rangle}}{2\sqrt3},
\frac{-a^{\langle23\rangle}-a^{\langle13\rangle}
+2a^{\langle12\rangle}}{6}
\right)$.
For simplicity, we fix the non-Abelian junction point 
$N_0$ at the origin,
namely we fix $a^{\langle12\rangle} = a^{\langle23\rangle} 
= a^{\langle13\rangle}$.
By using the real part of 
the $GL(\NC,{\bf C})$ transformations for 
the $V$-equivalence relation, 
we can set these to zero.
As a result we get 
$A_1= -a^{\langle14\rangle}
\left(-\frac{\sqrt3}{2},-\frac{1}{2}\right)$,
$A_2= -a^{\langle24\rangle}
\left(\frac{\sqrt3}{2},-\frac{1}{2}\right)$,
$A_3 = - a^{\langle34\rangle}\left(0,1\right)$ and
$N_0 = (0,0)$.
Therefore $L_*\equiv -a^{\langle *4\rangle}$ gives the 
length of the arm $N_0A_*$, 
 {with ``$*$" denoting $1,2,3$.}
These parameters have to satisfy the Pl\"ucker relation 
\be
e^{a^{\langle34\rangle}+ib^{\langle34\rangle}} - 
e^{a^{\langle24\rangle}+ib^{\langle24\rangle}} +
e^{a^{\langle14\rangle}+ib^{\langle14\rangle}} = 0,
\label{hex_pr}
\ee
where we fix NG mode coming from broken 
$U(1)_{\rm F}^3$ as
$b^{\langle12\rangle} = b^{\langle23\rangle} 
= b^{\langle13\rangle}=0$.
Notice that the overall phase of the Pl\"ucker relation 
can be absorbed by 
the imaginary part of 
the $GL(\NC,{\bf C})$ 
transformations for 
the $V$-equivalence relation.
At this stage the five parameters 
(position $N_0$ and $U(1)^3_{\rm F}$)
have been fixed, and there still remain two 
constraints of the Pl\"ucker relation and possibility of 
the imaginary part of 
the $GL(\NC,{\bf C})$ 
transformations for 
the $V$-equivalence relation. 
As a result, we obtain three independent parameters 
out of six parameters $a^{\langle *4\rangle}$ 
and $b^{\langle*4\rangle}$.  
The physical meaning of the Pl\"ucker relation 
can be understood as follows.
When the arms have comparable length 
$L_1\simeq L_2 \simeq L_3 >0$, the Pl\"ucker relation
determines phases $b^{\langle14\rangle},b^{\langle24\rangle}$ 
and $b^{\langle34\rangle}$ up to $V$-equivalence relation.
In this case, as we expected before, the shape of the web 
can be freely changed 
by changing the parameters $a^{\langle *4\rangle}$.
However, when one of the arms becomes extremely short, 
the situation drastically changes. 
For example, consider the situation where 
$L_2 = 0$ and $L_1 \simeq L_3 \gg 0$. 
Clearly such situation is forbidden by the Pl\"ucker relation
since the left hand side of Eq.(\ref{hex_pr}) is order one.
The Pl\"ucker relation only allows that two out of three 
arms are simultaneously
short, for example $L_1 \simeq L_2 \simeq 0$ while $L_3 \gg 0$.
This is a very big difference between the webs of $G_{4,2}$ and
the webs of ${\bf C}P^5$ 
 {(with the same positions of vertices in its web diagram)}. 
In the case of the webs of ${\bf C}P^5$
the lengths of all arms are completely independent 
because no Pl\"ucker relations exist. 
Furthermore, one can connect two anti-podal vertices as 
shown by a dashed line in Fig.~\ref{cp5}.
Contrary to this ${\bf C}P^5$ case, this kind of 
configuration is 
forbidden in the case of the webs of $G_{4,2}$, 
because such segment is not a single domain wall anymore 
(compare Fig.~\ref{model}(a) with Fig.~\ref{cp5}). 
This is precisely a physical consequence 
of the Pl\"ucker relation.
Now we can easily show that there exist the deformation
of the webs given in Fig.~\ref{model}(a2). 
For simplicity, we set
$a^{\langle 14\rangle} = a^{\langle 24\rangle}$
and fix $- a^{\langle 34\rangle}>0$ in the following. 
Depending on the value of $a^{\langle14\rangle}$ 
there are three cases 
with completely different shapes of the webs 
as shown in Fig.~\ref{hex_mod}(b): 
$a^{\langle14\rangle}/2 < 0$,
$0 < a^{\langle14\rangle}/2 < -a^{\langle34\rangle}$ and
$-a^{\langle34\rangle} < a^{\langle14\rangle}/2$, 
respectively.  
There exist three outer Abelian and 
one non-Abelian junctions in the first case,
one Abelian and one non-Abelian junctions and 
one intersection in the second case, and 
three outer non-Abelian and
one Abelian junctions in the third case.

\begin{figure}[ht]
\begin{center}
\begin{tabular}{ccc}
\includegraphics[height=4cm]{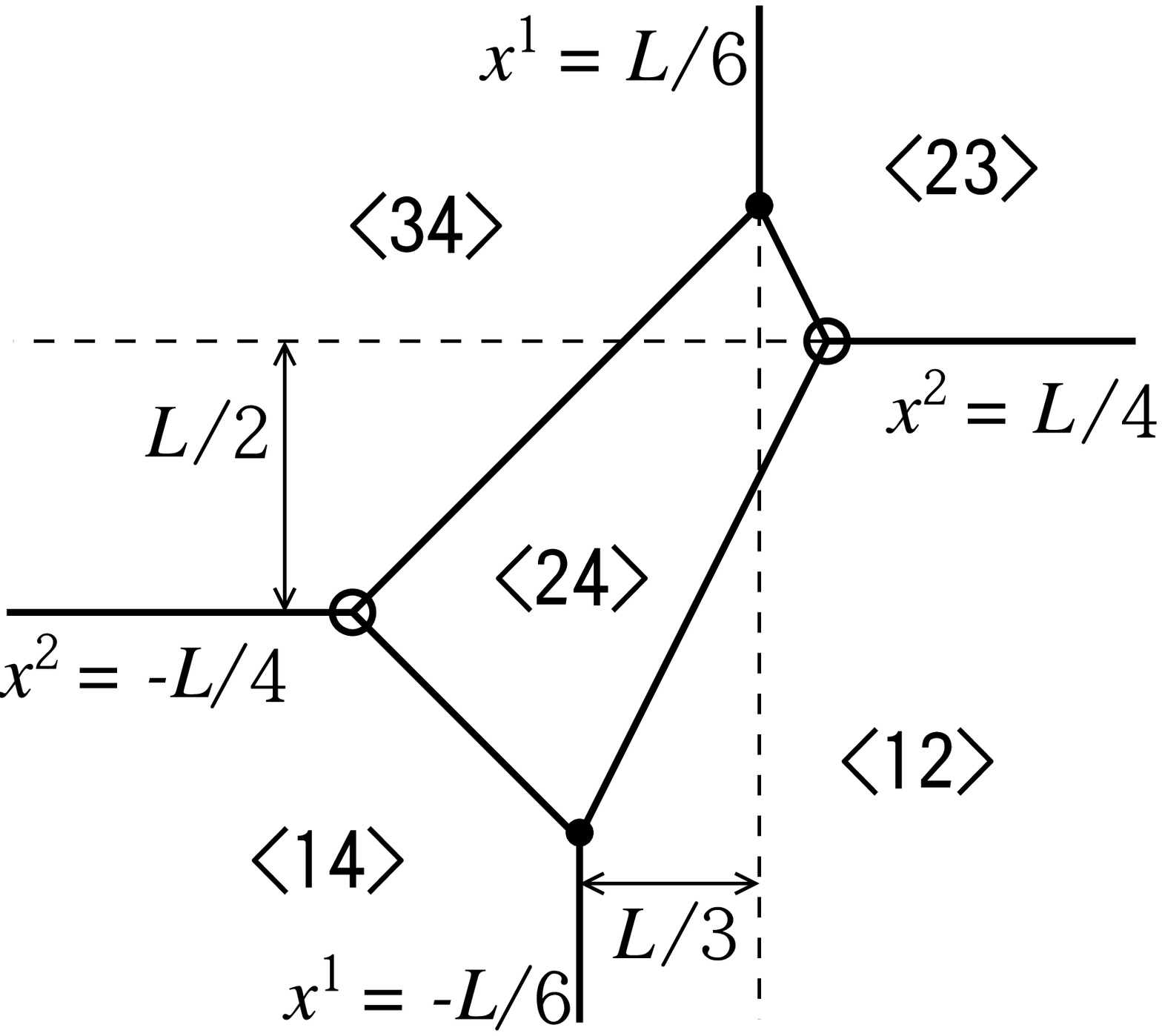}
&\qquad&
\includegraphics[height=4cm]{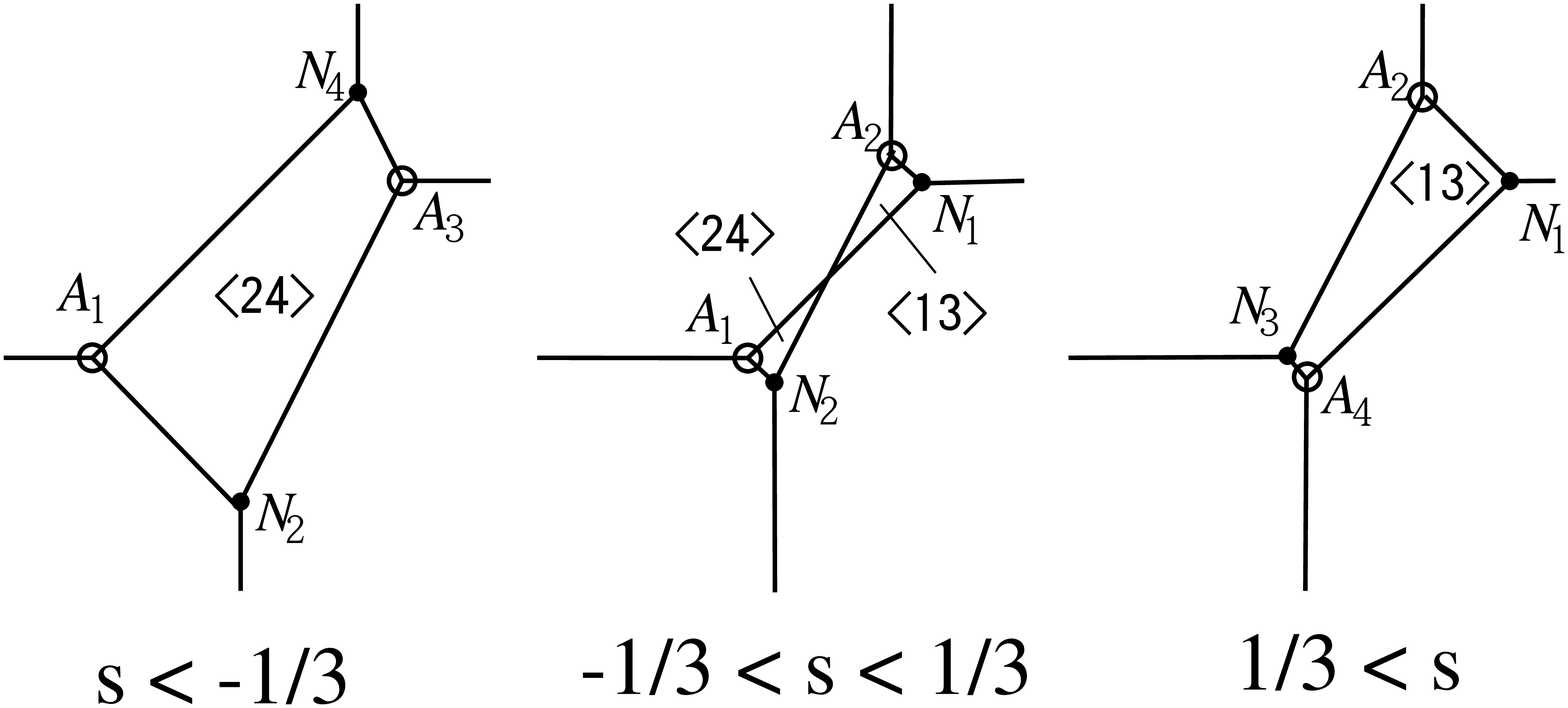}\\
\small{\sf (a) A configuration} & & 
\small{\sf (b) Deformation of the loop in the web}
\vspace*{-.3cm}
\end{tabular}
\caption{\small{\sf The web diagrams in the 
parallelogram-type mass arrangement.
Positions of the Abelian junction $A_*$ and 
the non-Abelian junction $N_*$ are given by
$A_1=(s-1,-1),
A_2 = (\frac{2}{3},\frac{4}{3}+s),
A_3 = (\frac{1-s}{2},1),
A_4= (-\frac{2}{3},-\frac{2}{3}-s),
N_1 = (1+s,1),
N_2= (-\frac{2}{3},s-\frac{4}{3}),
N_3= (-\frac{1+s}{2},-1),
N_4 = (\frac{2}{3},\frac{2}{3}-s)$ in unit of $\frac{L}{4}$.}}
\label{para}
\vspace*{-.3cm}
\end{center}
\end{figure}

Next let us examine the parallelogram-type web with 
the mass parameters $M = {\rm diag}(1/2-i,3/2,1/2+i,-3/2)$, 
like Fig.~\ref{model}(b). 
The web graphs have four external legs as shown in 
Fig.~\ref{para}.
A characteristic feature of these diagrams have 
loops unlike the hexagon-type web.
Here we concentrate on variation of shape of the web 
like Fig.~\ref{model}(b2).
The shapes of the loops are controlled by the weights 
$e^{{\cal W}^{\langle 13\rangle}}$ and 
$e^{{\cal W}^{\langle 24\rangle}}$ 
for the internal vacua $\langle 13\rangle$ 
and $\langle 24\rangle$. 
We start with the diagram in Fig.~\ref{para}(a).
To avoid inessential complications, we fix four external 
legs.
This can be done by setting 
$a^{\langle12\rangle} = a^{\langle34\rangle}$, 
$a^{\langle14\rangle} = a^{\langle23\rangle}$ 
and fixing $L \equiv a^{\langle12\rangle} 
+ a^{\langle 34\rangle}
- a^{\langle 14\rangle} - a^{\langle 23\rangle}$.
The parameter $L$ determines the size of the loop. 
By using real part of 
the $GL(\NC,{\bf C})$ 
transformations for 
the $V$-equivalence relation, we can set 
$a^{\langle 12\rangle} = L/4$. 
Then positions of the four external legs are 
given by $x^1=\pm L/6$ and $x^2 = \pm L/4$ as seen 
in Fig.~\ref{para}.
We also fix parameters 
$b^{\langle 12\rangle},b^{\langle 34\rangle}, 
b^{\langle 14\rangle}$
and $b^{\langle 23\rangle}$. 
Note that $U(1)_{\rm F}^3$ act as 
$\delta b^{\langle AB \rangle} = \theta_A + \theta_B$ with
$\sum_{A=1}^4 \theta_A=0$. 
Then the imaginary part of the
the $GL(\NC,{\bf C})$ 
transformations for 
the $V$-equivalence relation 
and two out of $U(1)^3_{\rm F}$ are fixed. 
Only one $U(1)_{\rm F}$ transformation 
defined by 
$\varphi/4 \equiv \theta_1=-\theta_2=\theta_3=-\theta_4$ 
remains under the above fixing. 
The remaining moduli parameters are
$a^{\langle13\rangle}+ib^{\langle13\rangle}= (u+v)L/4$ and 
$a^{\langle24\rangle}+ib^{\langle24\rangle}= (u-v)L/4$ 
$(u,v\in {\bf C})$
which control the shape of the loop. Notice that 
$u$ is invariant under the $\varphi$ transformation while
$v$ transforms as $(L/2)\delta_\varphi v = i\varphi$.
The imaginary part of $v$ is then the NG mode 
for the broken $U(1)_{\rm F}$ symmetry.
The Pl\"ucker relation determines the parameter $u$ as
\be
e^{\frac{L}{2}+i(b^{\langle12\rangle}+b^{\langle 34\rangle})} - 
e^{\frac{L}{2}u} + 
e^{-\frac{L}{2}+i(b^{\langle14\rangle}+b^{\langle 23\rangle})} = 0.
\ee
In the following we consider that the size of the loop is sufficiently large,
namely $L\gg1$. This implies that 
$u \simeq 1$ holds 
($u \simeq -1$ holds for $L\ll-1$).
At this stage the web given in Fig.~\ref{para}(b) is 
controlled by one moduli parameter 
$s\equiv {\rm Re}(v)$. 
In the parameter region $s<-1/3$, 
only the vacuum $\langle 24 \rangle$
arises in the loop as in the left figure. 
The web has two Abelian junctions $A_1$ and $A_3$ and
two non-Abelian junctions $N_2$ and $N_4$.
When $s=-1/3$ holds, 
the Abelian junction $A_3$ and the non-Abelian junction $N_4$ 
get together, 
and they pass through each other when $s$ becomes larger 
than $-1/3$. 
In the parameter region where $-1/3 < s < 1/3$, 
both the vacua $\langle 24 \rangle$
and $\langle 13 \rangle$ appear in the loop as in the 
second figure.
Here the web has two Abelian junctions $A_1$ and $A_2$ 
and two non-Abelian junctions
$N_1$ and $N_2$. 
When $s=1/3$ holds, the Abelian junction $A_1$ 
and the non-Abelian junction $N_2$ 
get together. 
When $s$ is larger than $1/3$, the vacuum 
$\langle 24 \rangle$ disappears
in the loop. 
In the parameter region $s>1/3$ 
the web has two Abelian junctions $A_2$ and $A_4$
and two non-Abelian junctions $N_1$ and $N_3$ as 
in the third figure. 

Notice that the value of the complex moduli parameter $v$ 
does not change the boundary condition of the web. 
Then we can consider the effective theory on the world 
volume of the web by promoting $v$ to a  
field $v(t,x^3)$ 
whose real part deforms the loop and 
imaginary part is the NG mode for the 
broken $U(1)_{\rm F}$. 
Such effective theory is 
a 1+1 dimensional ${\cal N}=(2,0)$ 
SUSY 
sigma model~\cite{Eto:2005cp,Eto:2005sw}.

%%%%%%%%%%%%%%%%%%%%%%%%%%%%%%%%%%%%%%%%%%%%%%%%%%%%%%%%%%%%%%%%%%%%%%%%%%%%%%%
\subsubsection*{Acknowledgements}

We would like to thank David Tong for useful discussions.
This work is supported in part by Grant-in-Aid for 
Scientific Research from the Ministry of Education, 
Culture, Sports, Science and Technology, Japan No.17540237 
(N.~S.) and 16028203 for the priority area ``origin of 
mass'' (N.~S.). 
The works of M.~N. and K.~O. (M.~E. and Y.~I.) are 
supported by Japan Society for the Promotion 
of Science under the Post-doctoral (Pre-doctoral) Research 
Program.   
M.~E., M.~N. and K.~O. wish to thank DAMTP and IPPP for 
their hospitality at the last stage of this work. 

%%%%%%%%%%%%%%%%%%%%%%%%%%%%%%%%%%%%%%%%%%%%%%%%%%%%%%%%%%%

%%%%%%%%%%%%%%%%%%%%%%%%%%%%%%%%%%%%%%%%%%%


\begin{thebibliography}{100}


%\cite{Isozumi:2004jc}
\bibitem{Isozumi:2004jc}
  Y.~Isozumi, M.~Nitta, K.~Ohashi and N.~Sakai,
  %``Construction of non-Abelian walls and their complete moduli space,''
  Phys.\ Rev.\ Lett.\  {\bf 93}, 161601 (2004)
  [arXiv:hep-th/0404198];
  %%CITATION = HEP-TH 0404198;%%
%
%\cite{Isozumi:2004va}
%\bibitem{Isozumi:2004va}
%  Y.~Isozumi, M.~Nitta, K.~Ohashi and N.~Sakai,
  %``Non-Abelian walls in supersymmetric gauge theories,''
  Phys.\ Rev.\ D {\bf 70}, 125014 (2004)
  [arXiv:hep-th/0405194];
  %%CITATION = HEP-TH 0405194;%%
%
%\cite{Isozumi:2004vg}
%\bibitem{Isozumi:2004vg}
%  Y.~Isozumi, M.~Nitta, K.~Ohashi and N.~Sakai,
  %``All exact solutions of a 1/4 Bogomol'nyi-Prasad-Sommerfield equation,''
  Phys.\ Rev.\ D {\bf 71}, 065018 (2005)
  [arXiv:hep-th/0405129].
  %%CITATION = HEP-TH 0405129;%%

%\cite{Eto:2004vy}
\bibitem{Eto:2004vy}
  M.~Eto, Y.~Isozumi, M.~Nitta, K.~Ohashi, K.~Ohta and N.~Sakai,
  %``D-brane construction for non-Abelian walls,''
  Phys.\ Rev.\ D {\bf 71}, 125006 (2005)
  [arXiv:hep-th/0412024].
  %%CITATION = HEP-TH 0412024;%%

%\cite{Eto:2005wf}
%\bibitem{Eto:2005wf}
%M.~Eto, Y.~Isozumi, M.~Nitta, K.~Ohashi, K.~Ohta, N.~Sakai and Y.~Tachikawa,
  %``Global structure of moduli space for BPS walls,''
% Phys.\ Rev.\ D {\bf 71}, 105009 (2005)
%  [arXiv:hep-th/0503033].
  %%CITATION = HEP-TH 0503033;%%

%\cite{Gibbons:1999np}
\bibitem{Gibbons:1999np}
  G.~W.~Gibbons and P.~K.~Townsend,
  %``A Bogomolnyi equation for intersecting domain walls,''
  Phys.\ Rev.\ Lett.\  {\bf 83}, 1727 (1999)
  [arXiv:hep-th/9905196];
  %%CITATION = HEP-TH 9905196;%%
%
%\cite{Carroll:1999wr}
%\bibitem{Carroll:1999wr}
  S.~M.~Carroll, S.~Hellerman and M.~Trodden,
  %``Domain wall junctions are 1/4-BPS states,''
  Phys.\ Rev.\ D {\bf 61}, 065001 (2000)
  [arXiv:hep-th/9905217].
  %%CITATION = HEP-TH 9905217;%%


%\cite{Eto:2005cp}
\bibitem{Eto:2005cp}
  M.~Eto, Y.~Isozumi, M.~Nitta, K.~Ohashi and N.~Sakai,
  ``Webs of Walls,''
  arXiv:hep-th/0506135.
  %%CITATION = HEP-TH 0506135;%%

%\cite{Kakimoto:2003zu}
\bibitem{Kakimoto:2003zu}
  K.~Kakimoto and N.~Sakai,
  %``Domain wall junction in N = 2 supersymmetric QED in four dimensions,''
  Phys.\ Rev.\ D {\bf 68}, 065005 (2003)
  [arXiv:hep-th/0306077].
  %%CITATION = HEP-TH 0306077;%%


%\cite{Oda:1999az}
\bibitem{Oda:1999az}
  H.~Oda, K.~Ito, M.~Naganuma and N.~Sakai,
  %``An exact solution of BPS domain wall junction,''
  Phys.\ Lett.\ B {\bf 471}, 140 (1999)
  [arXiv:hep-th/9910095];
  %%CITATION = HEP-TH 9910095;%%
%\cite{Ito:2000zf}
%\bibitem{Ito:2000zf}
  K.~Ito, M.~Naganuma, H.~Oda and N.~Sakai,
  %``Nonnormalizable zero modes on BPS junctions,''
  Nucl.\ Phys.\ B {\bf 586}, 231 (2000)
  [arXiv:hep-th/0004188];
  %%CITATION = HEP-TH 0004188;%%
%\cite{Ito:2000mt}
%\bibitem{Ito:2000mt}
%  K.~Ito, M.~Naganuma, H.~Oda and N.~Sakai,
  %``An exact solution of BPS junctions and its properties,''
  Nucl.\ Phys.\ Proc.\ Suppl.\  {\bf 101}, 304 (2001)
  [arXiv:hep-th/0012182];
  %%CITATION = HEP-TH 0012182;%%  
%\cite{Shifman:1999ri}
%\bibitem{Shifman:1999ri}
  M.~A.~Shifman and T.~ter Veldhuis,
  %``Calculating the tension of domain wall junctions and vortices in
  %generalized Wess-Zumino models,''
  Phys.\ Rev.\ D {\bf 62}, 065004 (2000)
  [arXiv:hep-th/9912162];
  %%CITATION = HEP-TH 9912162;%%
%\cite{Naganuma:2001br}
%\bibitem{Naganuma:2001br}
  M.~Naganuma, M.~Nitta and N.~Sakai,
  %``BPS walls and junctions in SUSY nonlinear sigma models,''
  Phys.\ Rev.\ D {\bf 65}, 045016 (2002)
  [arXiv:hep-th/0108179]. 
%; 
  %%CITATION = HEP-TH 0108179;%%
%\cite{Naganuma:2002su}
%\bibitem{Naganuma:2002su}
%M.~Naganuma, M.~Nitta and N.~Sakai,
%``BPS walls and junctions in N = 1 SUSY nonlinear sigma models,''
%Proceedings of 3rd International Sakharov Conference On Physics, 
%edited by A. Semikhatov {\it et al.} (Scientific World Pub., 2003) 
%p.537 - p.549, %24-29 Jun 2002, Moscow, Russia [arXiv:hep-th/0210205].
%%CITATION = HEP-TH 0210205;%%


%\cite{Eto:2005sw}
\bibitem{Eto:2005sw}
M.~Eto, Y.~Isozumi, M.~Nitta and K.~Ohashi,
``1/2, 1/4 and 1/8 BPS equations in SUSY Yang-Mills-Higgs systems: Field
 theoretical brane configurations,''
 arXiv:hep-th/0506257.
  %%CITATION = HEP-TH 0506257;%%


%\cite{Arai:2003tc}
\bibitem{Arai:2003tc}
  M.~Arai, M.~Nitta and N.~Sakai,
  %``Vacua of massive hyper-Kaehler sigma models of non-Abelian quotient,''
  Prog.\ Theor.\ Phys.\  {\bf 113}, 657 (2005)
  [arXiv:hep-th/0307274].
  %%CITATION = HEP-TH 0307274;%%  

%\cite{Shifman:2002jm}
\bibitem{Shifman:2002jm}
M.~Shifman and A.~Yung,
%``Domain walls and flux tubes in N = 2 SQCD: D-brane prototypes,''
Phys.\ Rev.\ D {\bf 67}, 125007 (2003)
[arXiv:hep-th/0212293].
%%CITATION = HEP-TH 0212293;%% 

%\cite{Gorsky:1999hk}
\bibitem{Gorsky:1999hk}
  A.~Gorsky and M.~A.~Shifman,
  %``More on the tensorial central charges in N = 1 supersymmetric gauge
  %theories (BPS wall junctions and strings),''
  Phys.\ Rev.\ D {\bf 61}, 085001 (2000)
  [arXiv:hep-th/9909015].
  %%CITATION = HEP-TH 9909015;%%

%\cite{Chibisov:1997rc}
\bibitem{Chibisov:1997rc}
  B.~Chibisov and M.~A.~Shifman,
  %``BPS-saturated walls in supersymmetric theories,''
  Phys.\ Rev.\ D {\bf 56}, 7990 (1997)
  [Erratum-ibid.\ D {\bf 58}, 109901 (1998)]
  [arXiv:hep-th/9706141].
  %%CITATION = HEP-TH 9706141;%%



%\cite{Cherkis:2000cj}
\bibitem{Cherkis:2000cj}
  S.~A.~Cherkis and A.~Kapustin,
%``Nahm transform for periodic monopoles and N = 2 super Yang-Mills  theory,''
  Commun.\ Math.\ Phys.\  {\bf 218}, 333 (2001)
  [arXiv:hep-th/0006050].
  %%CITATION = HEP-TH 0006050;%%
  
\end{thebibliography}
\end{document}